\documentclass[aps,prd,numbers,amsmath,amssymb,nofootinbib,superscriptaddress,floatfix,showkeys,twocolumn,longbibliography,
]{revtex4-1}
\usepackage{graphicx,color,multirow}
\usepackage[latin1]{inputenc}
\usepackage{amsmath,amssymb}
\usepackage{epstopdf}
\usepackage{slashed}
\usepackage{soul}
\usepackage{amssymb}
\usepackage[normalem]{ulem}
\usepackage[caption=false]{subfig}

\newcommand{\lsim}{\mathrel{\mathop{\kern 0pt \rlap
  {\raise.2ex\hbox{$<$}}}
  \lower.9ex\hbox{\kern-.190em $\sim$}}}
\newcommand{\gsim}{\mathrel{\mathop{\kern 0pt \rlap
  {\raise.2ex\hbox{$>$}}}
  \lower.9ex\hbox{\kern-.190em $\sim$}}}

\graphicspath{{figures/}}

\def  \bcen   {\begin{center}}
\def  \ecen   {\end{center}}
\def  \beq    {\begin{equation}}
\def  \eeq    {\end{equation}}
\def  \beqa   {\begin{eqnarray}}
\def  \eeqa   {\end{eqnarray}}

\def\bea{\begin{eqnarray}}
\def\eea{\end{eqnarray}}
\begin{document}

\title{
Evidence for a $\sim 43$ GeV $\gamma$-ray line signal in a stacking analysis of the Virgo, Fornax, and Ophiuchus Galaxy clusters
}

\author{Yi-Zhong Fan}
\email{yzfan@pmo.ac.cn}
\affiliation{Key Laboratory of Dark Matter and Space Astronomy, Purple Mountain Observatory, Chinese Academy of Sciences, Nanjing 210023, China}
\affiliation{School of Astronomy and Space Science, University of Science and Technology of China, Hefei, Anhui 230026, China}
\author{Zhao-Qiang Shen}
\email{zqshen@pmo.ac.cn}
\affiliation{Key Laboratory of Dark Matter and Space Astronomy, Purple Mountain Observatory, Chinese Academy of Sciences, Nanjing 210023, China}
\author{Yun-Feng Liang}
\email{liangyf@gxu.edu.cn}
\affiliation{Guangxi Key Laboratory for Relativistic Astrophysics, School of Physical Science and Technology, Guangxi University, Nanning 530004, China}
\author{Xiang Li}
\email{xiangli@pmo.ac.cn}
\affiliation{Key Laboratory of Dark Matter and Space Astronomy, Purple Mountain Observatory, Chinese Academy of Sciences, Nanjing 210023, China}
\affiliation{School of Astronomy and Space Science, University of Science and Technology of China, Hefei, Anhui 230026, China}
\author{Kai-Kai~Duan}
\author{Zi-Qing~Xia}
\affiliation{Key Laboratory of Dark Matter and Space Astronomy, Purple Mountain Observatory, Chinese Academy of Sciences, Nanjing 210023, China}
\author{Xiao-Yuan Huang}
\author{Lei Feng}
\author{Qiang Yuan}
\affiliation{Key Laboratory of Dark Matter and Space Astronomy, Purple Mountain Observatory, Chinese Academy of Sciences, Nanjing 210023, China}
\affiliation{School of Astronomy and Space Science, University of Science and Technology of China, Hefei, Anhui 230026, China}
\date{\today}

\begin{abstract}
As the largest gravitationally bound objects in the Universe, galaxy clusters have provided the first piece of evidence for the presence of dark matter and may be suitable targets for indirect dark matter searches. Among various signals, the GeV-TeV $\gamma$-ray line has been taken as the smoking-gun signal of the dark matter annihilation and decay since no known astrophysical or physical process(es) could generate such a peculiar spectrum. With 15.5 years of Fermi-LAT P8R3 publicly available data, we search for the $\gamma$-ray line emission in the directions of 13 massive galaxy clusters at redshifts $z \leq 0.028$ with an unbinned likelihood analysis.
A $\gamma$-ray line signal at $\sim 43.2$ GeV has a net test statistic (TS) value of $\approx 30$ if we take into account only the data in the directions of the Virgo, Fornax, and Ophiuchus clusters, three massive clusters with the highest J-factors expected to generate the dark matter annihilation signal. The signal still presents when the data of 10 other nearby massive clusters have also been included, though the TS value decreases to $\approx 21$, likely because of their lower signal-to-noise ratios. The absence of this signal in the inner Galaxy disfavors both the instrumental effect and the canonical dark matter annihilation interpretation, and a more sophisticated dark matter model or very peculiar astrophysical scenario might be needed. This $\gamma$-ray line signal, if intrinsic, could be unambiguously verified by the Very Large Area $\gamma$-ray Space Telescope in its first two years of performance.  
\end{abstract} 

\maketitle
{\it Introduction---} The observations of various gravitational phenomena at different scales strongly suggest the presence of dark matter (DM) \cite{Bertone:2016nfn}. Many interesting dark matter particle candidates have been proposed in the literature, and the weakly interacting massive particles (WIMPs) are the leading ones since they can naturally explain today's relic density of DM \cite{Feng:2010gw}. The WIMPs may annihilate each other or decay, and finally, generate stable standard model particle pairs, such as gamma rays, neutrinos and antineutrinos, electrons and positrons, protons and antiprotons, and so on. The successful identification of such signals, in turn, can be used to infer the properties of the dark matter particles.  Though dedicated efforts have been made in past decades, no definitive dark matter signal has been identified in either gamma rays  or cosmic rays \cite{Fermi-LAT:2016afa,Conrad:2017pms,Fan:2022dck,HESS:2022ygk,MAGIC:2022acl,LHAASO:2024upb}. One challenge is that the resulting energy spectra of the products are usually smooth, so it is hard to distinguish them from astrophysical backgrounds \cite{Feng:2010gw,Fermi-LAT:2016afa}. The situation could change if the products have a sharp structure like a line, as expected in the annihilation or decay of WIMPs $\chi$ into a two-body final state $\gamma \chi$ \cite{Bergstrom:1988} (or the decay of the gravitinos through $\chi\rightarrow \gamma \nu$ \cite{Ibarra:2007wg}). The line energy is $E_\gamma=m_\chi(1-m_X^{2}/4m_\chi^{2})$ for dark matter annihilation, where $X$ could be either $\gamma$, $Z_0$ or $h_0$, depending on the mass of the dark matter particle \cite{Bergstrom:1988,Rudaz:1991}.  

Shortly after the first release of the Fermi-LAT data, quite a few groups searched for $\gamma$-ray line signals in the direction of the inner Galaxy and did find evidence for the presence of a $\sim 130$ GeV line \cite{Bringmann:2012vr,Weniger:2012tx,Tempel:2012ey,Su:2012ft}. Such a signal, however, turns out to be a systematic effect of the early data reconstruction of Fermi-LAT \cite{Fermi-LAT:2015kyq,2019PhRvD..99l3519L,DeLaTorreLuque:2023fyg}. This null result has also been confirmed by the independent observation of the DArk Matter Particle Explorer (DAMPE) \cite{DAMPE:2021hsz,Cheng:2023chi}. 
With the Pass 8 data of Fermi-LAT, we have concentrated on the 
galaxy clusters, the most massive gravitational bound systems in the Universe, which contain a large number of substructures that may be promising targets for dark matter search \cite{Gao:2011rf,Fermi2010JCAP,Anderson:2015dpc,Huang2012JCAP,Lisanti2018PhRvL,ThorpeMorgan2021MNRAS,DiMauro2023PhRvD,Li:2025vek}, and found hints for a $\sim 43$ GeV $\gamma$-ray line in the direction of a group of nearby massive galaxy clusters \cite{Liang:2016pvm,Shen:2021fie}. 
In this work, we show that this signal still presents in the latest data and its significance, though characterized by a sizable drop around MJD 57500, grows again at late times. 
If intrinsic, this signal will be unambiguously verified by the Very Large Area $\gamma$-ray Space Telescope (VLAST),  a MeV-TeV detector distinguished by a peak acceptance of $\sim 12~{\rm m^{2}~sr}$ and an excellent energy resolution of $\sim 1.3\%$ at 50 GeV, which has been proposed by some DAMPE people \cite{{Fan:2022}},  in $\sim 1-2$ yr of performance.

{\it Methodology---}
Our baseline sample consists of 13 galaxy clusters with large J-factors. Most were selected from the extended HIFLUGCS catalog \cite{Reiprich:2001zv,Chen:2007sz} and are basically the same as that used in Liang et al. \cite{Liang:2016pvm}, except that we removed Perseus because of the strong gamma-ray activity of the radio galaxy NGC~1275 in the cluster~\citep{Baghmanyan2017,Cheng2021,2024MNRAS.tmp.2493C}, and removed 3C129 as well as A3627 for their low Galactic latitudes and high background. 
The Ophiuchus cluster is also at a low latitude and therefore is neglected in some previous works due to the strong background, but its J-factor is large, and the expected signal-to-noise ratio is still reasonably high, which is helpful in the DM search.
The region of interest (ROI) for each galaxy cluster is defined with a circular region located at the cluster center with an angular radius of $\theta_{\rm ROI,0} = \theta_{\rm 200,c} \equiv \arctan(R_{\rm 200,c}/d_{\rm A})$, where $d_{\rm A}$ is the angular diameter distance and $R_{\rm 200,c}$ is the central value of the virial radius.
We test the effect of the point spread function by using the ROI radii of $\theta_{\rm 200,c}+r_{68}$ and find that the results change only slightly, so we keep the current ROI definition.
In our analysis, the overlapped region of Virgo and M49 is taken into account only once.
Please refer to Table~\ref{tab:appx:clusters} of the Supplemental Material~\citep{supp} for a summary of our sample.

We take the Fermi-LAT P8R3\_V3 data based on the most recent event-level analysis, which alleviates the background cosmic rays leaked from the ribbons of the anticoincidence detector \cite{2018arXiv181011394B}. The ULTRACLEAN events are adopted to reduce the cosmic-ray contamination \cite{Ackermann:2012}. 
We take into account only the EDISP(1+2+3) data ({\tt evtype=896}), collected between 2008 October 27 and 2024 May 2 (Fermi Mission Elapsing Time between 246823875 and 736304518), since the energy resolution of the EDISP0 data is significantly worse than the other events \cite{Fermi-LAT:2013jgq}. 
We further select the events with zenith angles less than $90^\circ$ and apply the quality filter cut {\tt (DATA\_QUAL==1)\&\&(LAT\_CONFIG==1)}.
Such a selection removes the Earth limb events and the observations during the bright gamma-ray burst GRB~221009A and strong solar flares.\footnote{
GRB~120526A and GRB~201208A appeared within the ROIs of the Fornax and A1060, respectively.
We perform a test by excluding the time intervals of the bursts and find the results barely change, so we keep the quality filter above.
}
Then, the photons within the ROIs of the clusters are collected.
The {\sc Fermitools v2.2.0} are utilized in the analyses.

We perform an unbinned likelihood analysis, employing a sliding-window technique and splitting
the analysis into a series of energy windows, to search for spectral lines \cite{Liang:2016pvm,Fermi-LAT:2015kyq,Ackermann:2012,Weniger:2012tx,Bringmann:2012vr}.
We consider a series of line energies $E_{\rm line}$ from 5~GeV to 300~GeV with the difference between two adjacent energies being $0.5\sigma_E$, where $\sigma_E$ is the half width of the $68\%$ containment energy dispersion for EDISP(1+2+3) events.
Each window covers the energy range of $[0.5 E_{\rm line}, 1.5 E_{\rm line}]$, which can well balance the statistical and systematic uncertainties~\citep{Fermi-LAT:2015kyq}.
The step size of the line energies and the energy window width were also adopted by the Fermi-LAT team in~\citep{Fermi-LAT:2015kyq}.
Within each window, the unbinned likelihood fit is carried out by assuming a power-law background. 
The small width of the energy window ensures the power-law function as a good approximation to the background spectrum.
The observational data toward the galaxy clusters are fitted with both null (purely power law) and signal (power law plus a line component) models.

The likelihood for the null model is expressed as
\begin{equation}
\label{eq:lnLnull}
\begin{aligned}
{\rm ln}{\mathcal L}_{\rm null}(\Theta_{\rm b}) = \sum_{i = 1}^{N_{\rm ph}} {\rm ln} \left[F_{\rm b}(E_{i};\Theta_{\rm b}) \bar{\epsilon}(E_{i})\right] \\
- \int F_{\rm b}(E;\Theta_{\rm b}) \bar{\epsilon}(E) {\rm d}E,
\end{aligned}
\end{equation}
where
$N_{\rm ph}$ is the number of photons in the analysis,
$E_{i}$ is the energy of each Fermi-LAT photon within ROI, $F_{\rm b}(E)=N_{\rm b}\, (E/{\rm 2~GeV})^\Gamma$ is the power-law background and $\Theta_{\rm b}=\{ N_{\rm b}, \Gamma\}$ represents the nuisance parameters of the background.
For the signal model, the likelihood is
\begin{equation}
\label{eq:lnLsig}
\begin{aligned}
{\rm ln}{\mathcal L}_{\rm sig}&(N_{\rm s},E_{\rm line},\Theta_{\rm b}) \\
& = \sum_{i = 1}^{N_{\rm ph}} {\rm ln} [F_{\rm b}(E_{i};\Theta_{\rm b}) \bar{\epsilon}(E_{i}) + F_{\rm s}(E_{i})\bar{\epsilon}(E'_{\rm line})] \\
& - \int [F_{\rm b}(E;\Theta_{\rm b}) \bar{\epsilon}(E) + F_{\rm s}(E) \bar{\epsilon}(E'_{\rm line})] {\rm d}E,
\end{aligned}
\end{equation}
with $F_{\rm s}(E) = N_{\rm s} \bar{D}(E;E'_{\rm line})$ the line component at an observed line energy of $E'_{\rm line}$ and $\bar{D}(E;E'_{\rm line})$ the exposure-weighted average of the energy dispersion function. The $\bar{\epsilon}$ is the solid angle weighted average of the exposures of the clusters.
For details of the calculation of the exposure $\bar{\epsilon}(E)$ and the energy dispersion $\bar{D}(E;E'_{\rm line})$, please see Sec.~\ref{appx:likelihood} of the Supplemental Material~\citep{supp}, as well as Ref.~\cite{Liang:2016pvm,Ackermann:2012}. 
To improve the line search sensitivity, we take into account the redshifts of the clusters by using $E'_{\rm line}=E_{\rm line}/(1+z)$, where $E_{\rm line}$ is the source-frame energy of the line produced in the cluster at the redshift of $z$.
During the analysis, the minimization of the likelihood is implemented by the {\sc Python} package {\sc iminuit}~\citep{Minuit1975,iminuit2020}.

A test statistic (TS) is defined as ${\rm TS} \equiv 2 \ln ({\hat {\mathcal L}}_{\rm sig}/{\hat{\mathcal L}}_{\rm null})$, where $\hat{\mathcal L}$ is the best-fit likelihood value.
The local significance of a line signal can be approximated as the square root of the TS value. A more accurate, realistic estimation of the global significance will be obtained through Monte Carlo (MC) simulations (see Sec.~\ref{appx:trials} of the Supplemental Material~\citep{supp}).

\begin{figure}
	\centering  
    \includegraphics[width=0.48\textwidth]{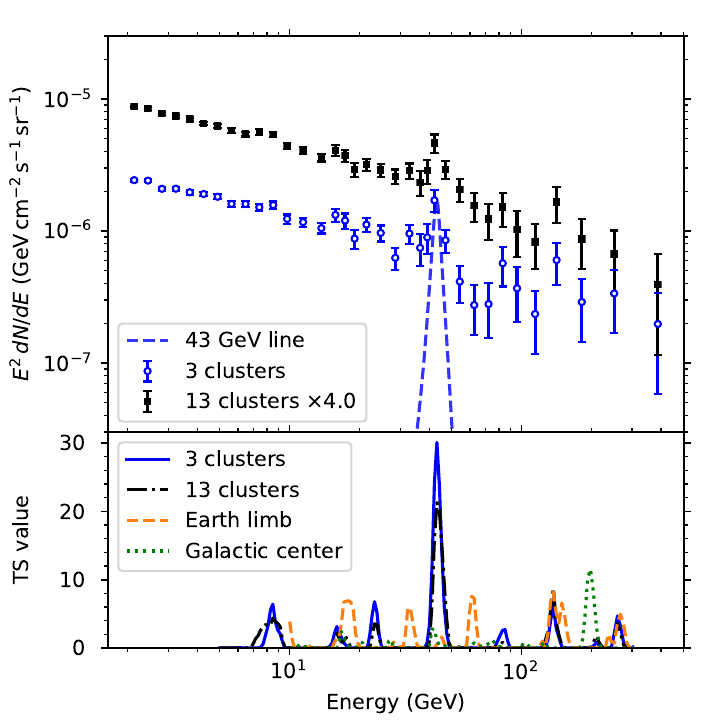}
\caption{The upper panel is for the spectral energy distributions (SEDs) of the three galaxy clusters with the largest J-factors (the blue points) and the whole sample (the black points). The lower panel shows the corresponding test statistic values of the possible $\gamma$-ray line signal in the directions of the first three galaxy clusters (the blue line), the whole sample (the dotted-dashed black line), as well as the null result from the Earth limb
(the dashed orange line) and the $3^\circ$ region around the Galactic center (green dotted line).
}
\label{fig:SED}
\end{figure}

\begin{figure*}[!htb]
	\centering  
     \includegraphics[width=0.48\textwidth]{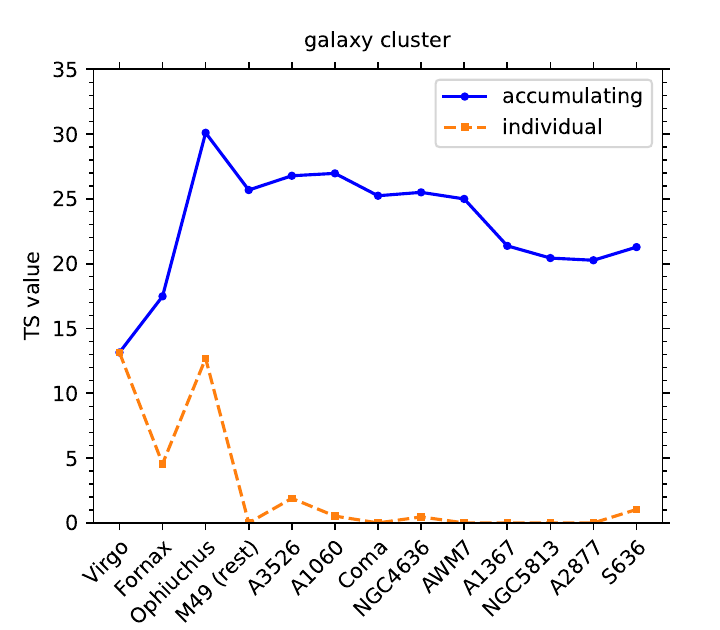}
     \includegraphics[width=0.48\textwidth]{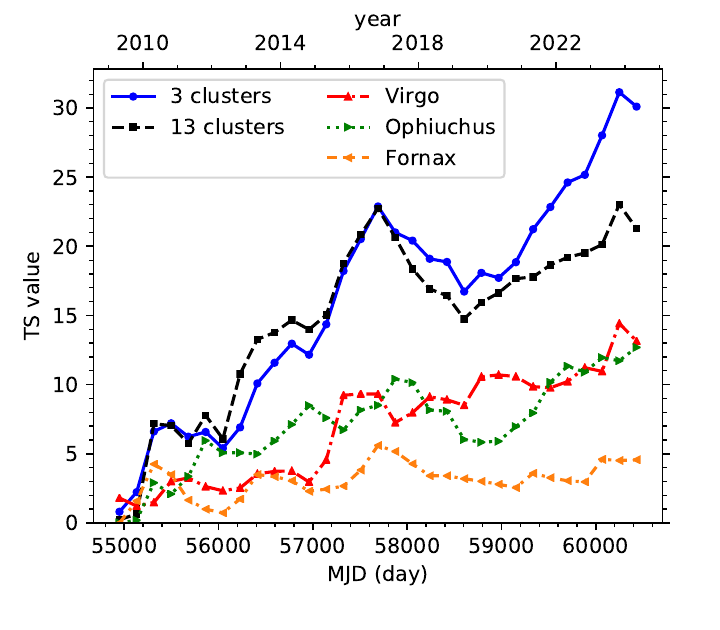}
\caption{The left panel is for the TS value of the line signal in the direction of each galaxy cluster (the orange line) and the evolution of the net TS value of a group of sources with the accumulation of sources (the blue line).
The right panel is for the time evolution of the TS values of the three galaxy clusters (the blue line), the whole sample (the black dashed line) as well as the individual clusters of Virgo (the red dot-dashed line), Ophiuchus (the green dotted line) and Fornax (the orange dot-dot-dashed line).   
}
\label{fig:TS-evolution}
\end{figure*}

{\it Results and discussion---}
Figure \ref{fig:SED} presents the spectral properties of the stacked galaxy clusters. The upper panel is for the spectral energy distributions (SEDs) of the three galaxy clusters with the largest J-factors (the data points in blue)  and the whole sample (the data points in black). The SED is defined as $\left({\rm d}N/{\rm d}E \right)_j=\sum_{i=1}^{N_{\rm gcl}} n_{ij}/(\bar{\epsilon}_j \Delta E_j \sum_{i=1}^{N_{\rm gcl}} \Omega_i)$, where $N_{\rm gcl}$ is the number of clusters, $\Omega_i$ is the solid angle and $n_{ij}$ is the photon number at the energy bin $E_j$ for the $i$-th cluster.
Clearly, a distinct spike displays in both spectra, though for the whole sample the signal is relatively weaker. This is likely attributed to its higher background in comparison to the case of the first three galaxy clusters. 

We then carry out the unbinned likelihood line search (i.e., the sliding-window data analysis). The results are reported in the lower panel of Figure \ref{fig:SED} and the TS value reaches $30.1$ ($21.3$) for the first three galaxy clusters (the whole sample).
The derivation of the global significance via the random sky simulations is detailed in Sec.~\ref{appx:trials} of the Supplemental Material~\citep{supp}.
It turns out that for the signal in the whole sample the global significance is about $3.7\sigma$, while for the three galaxy clusters with the largest J-factors the global significance is about $4.3\sigma$.
We have also carried out the same line search procedure for the Earth limb data above 10~GeV whose zenith angles are within $111^\circ-113^\circ$ and rocking angles are $>52^\circ$~\citep{2013PhRvD..88h2002A,Fermi-LAT:2015kyq,Liang:2016pvm}, and found a null result (see the dashed orange line in the lower panel of Figure~\ref{fig:SED} and Figure~\ref{fig:appx:earthlimb} of the Supplemental Material~\citep{supp}).
We have searched for lines in the inner $3^\circ$ region around the Galactic center and found no signal as well (green dotted line).
The line signal presents in various event classes and event types, and is also robust against the sliding-window size and background spectral model (see Sec.~\ref{appx:systematics} of the Supplemental Material~\citep{supp}).
Together with the nondetection in the inner Galaxy data \cite{Cheng:2023chi}, we find no evidence for a systematic error origin of our signal.

Figure \ref{fig:TS-evolution} provides the details of the TS evolution of the signal. In the left panel, we present the TS value of the line signal in the direction of each ROI and the evolution of the net TS value of a group of sources as the sources accumulate. The order of the galaxy clusters is based on the central values of their J-factors (without considering the possible boost factor, see Table \ref{tab:appx:clusters} in the Supplemental Material~\citep{supp}). 
No signal is detected in the direction of M49. 
Its J-factor may be smaller than the expected value considering the large uncertainty (see Sec.~\ref{appx:consistency} of the Supplemental Material~\citep{supp}).
This also likely indicates a small boost factor because the dark matter distribution substructures may be atypical in view of that 
M49 may have even already passed through Virgo once, as revealed by the extended X-ray study \cite{2019AJ....158....6S}. 
The right panel of Figure \ref{fig:TS-evolution} displays the time evolution of the TS values of the three galaxy clusters (the blue line) as well as the whole sample (the black dashed line). For the former, the net TS value roughly increases with time, though strong fluctuation is displayed in the time interval of MJD $57000-59000$, while for the whole sample, the net TS value peaks at about MJD 57700. After that there is a quick drop in the TS value. Since MJD 58600, the TS value has been increasing again. Roughly, the TS values of the signal in the directions of Virgo and Ophiuchus increase with time linearly, indicating a stable emission process.

\begin{figure}[ht!]
\centering  
\includegraphics[width=0.48\textwidth]{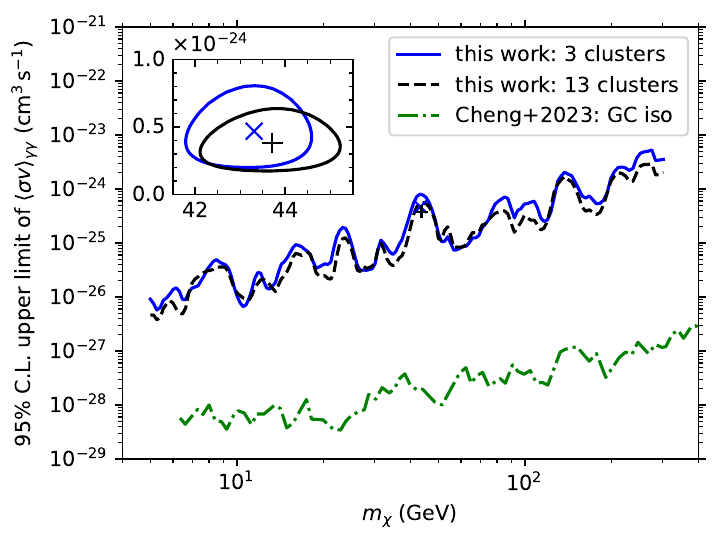}
\caption{The inset shows the best-fit
parameters needed in the dark matter annihilation origin of the $\sim 43$ GeV signal; note that the boost factors of the galaxy clusters have not been taken into account. At other energies, we present upper limits. The upper limits set by the null signals in the inner Galaxy \cite{Cheng:2023chi} are shown for comparison.}
\label{fig:Constraint}
\end{figure}

{\it (i) Stringent constraints on the canonical dark matter origin of the  $\gamma$-ray line signal---} It is widely believed that the inner Galaxy would have the brightest dark matter annihilation signal. In the canonical dark matter model, strong line emission would be expected from the inner Galaxy if the signals in the directions of the galaxy clusters are intrinsic. The null results, as widely reported \cite{Fermi-LAT:2015kyq,Cheng:2023chi}, are against the instrumental effect origin of the 43 GeV $\gamma$-ray line signal. The stringent upper limits are reported in Figure \ref{fig:Constraint} (see Sec.~\ref{appx:containment} of the Supplemental Material~\citep{supp} for more discussion on the constraints).
Clearly, the nondetection of the $\sim 43$ GeV line signal in the inner Galaxy region is strongly in tension with that of the galaxy clusters. 
There are some possible solutions. One is that the substructures of the galaxy clusters are so rich that their boost factors of the dark matter annihilation are high up to $\sim 10^{3}$. Such a possibility, though believed possible in some literature \cite{Gao:2011rf}, is not supported by the more recent numerical simulations \cite{Sanchez-Conde:2013yxa,Moline:2016pbm,Ishiyama:2019hmh}. 
The other possibility is that the canonical model \cite{2016PhRvD..94d3535F} is too simple and the real situation is much more complicated. In a recent systematic study of velocity-dependent dark matter annihilation (including the Sommerfeld enhancement) in a variety of astrophysical objects, it was found that for $s$-wave on resonances and for $p$-wave in the no-Sommerfeld enhancement regime the galaxy clusters can outshine all other classes of targets \cite{Lacroix:2022cjm}.
The nondetection of the continuum emission component in the direction of Virgo imposes an additional challenge (see Sec.~\ref{appx:continuum} of the Supplemental Material~\citep{supp}).
Note that a $2.5-3.0\sigma$ excess was reported in the continuous spectrum of clusters~\citep{DiMauro2023PhRvD}, suggesting the annihilation of $\sim 50~\rm GeV$ DM with a cross section $\sim 10$ times larger than ours. While it may have an underlying connection with the line signal reported in this Letter, more data are needed to clarify such an excess.
Another possibility is that the $\sim 43$ GeV line signal has an astrophysical origin, and hence, is irrelevant to the dark matter. 
One model, initially motivated by the 130 GeV signal reported in the inner Galaxy, is that the Comptonization of a cold, ultrarelativistic but mono-energetic electron-positron pulsar wind in the deep Klein-Nishina regime can yield narrow (with a width $\leq 0.1 E_{\rm line}$, Sec.~\ref{appx:linefeatures} in the Supplemental Material~\citep{supp}) distinct $\gamma$-ray line features \cite{2012A&A...547A.114A}. Though interesting, it seems challenging to realize such a specific scenario in the galaxy clusters. 

\begin{figure}
	\centering  
     \includegraphics[width=0.48\textwidth]{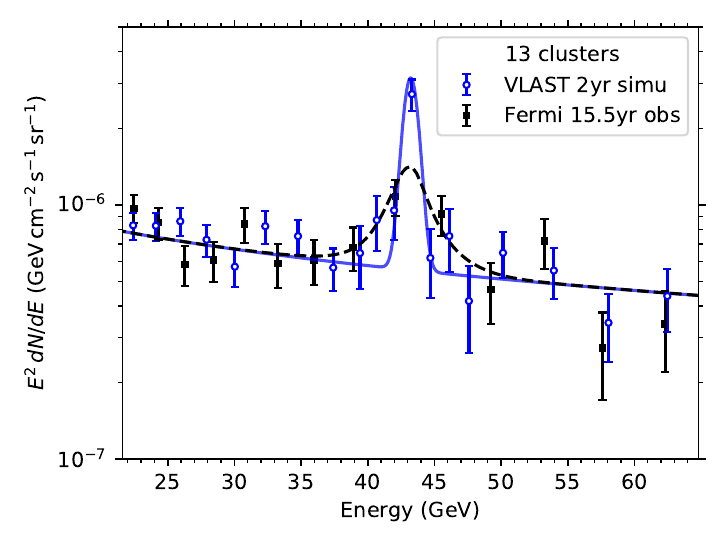}
\caption{The SED of the 2-yr VLAST simulation of the 13 galaxy clusters (the blue points) in the energy window around 43~GeV.
The blue solid line is the input model based on the best-fit spectral parameters of the Fermi-LAT observation.
The black points and the black dashed line are the Fermi-LAT SED and the fitted spectrum, respectively.
}
\label{fig:vlastsimu}
\end{figure}

{\it (ii) Prospect of testing the $\sim 43$ GeV $\gamma$-ray line signal with VLAST---} Fermi-LAT is in good condition and the continuous excellent performance provides the guarantee of collecting more high-quality data, which will be crucial to test whether the $\sim 43$ GeV signal is real or not. Nevertheless, the double of the current data can be achievable only until 2040. 
Furthermore, the ground based $\gamma$-ray telescopes, such as the Cherenkov Telescope Array~\citep{CTA2024_line}, suffer from significant cosmic-ray background and poor energy resolution at $\sim 40~\rm GeV$, making it nearly impossible to confirm the signal (see Sec.~\ref{appx:CTA} of the Supplemental Material~\citep{supp}).
Fortunately, the next generation $\gamma$-ray space telescopes are in proposal. VLAST, a Chinese space mission being proposed, is characterized by a large acceptance of $\sim 12~{\rm m^{2}~sr}$ at GeV-TeV energies and an excellent energy resolution of $\sim 1.3\%$ at 50 GeV \cite{Fan:2022}. Given the high acceptance and energy-resolution, such a telescope will be wonderful to clarify whether there is a distinct $\gamma$-ray line from the galaxy clusters.
We have simulated 2-yr VLAST observations of the 13 galaxy clusters (see Figure~\ref{fig:vlastsimu}), adopting the best-fit results of the signal as well as the background for the current Fermi-LAT data as the input, where the VLAST instrumental response functions are taken from~\citep{Fan:2022,Pan:2024adp}.
The TS value of the 43~GeV line is expected to be $\sim 73$.
Therefore, the successful performance of VLAST in $\sim 1-2$ years will be sufficient to confirm the presence of the line signal if the current Fermi-LAT signal is intrinsic.

{\it Conclusion---} We have analyzed the 15.5-yr observation data of Fermi-LAT in the directions of nearby very-massive galaxy clusters. With an unbinned likelihood approach, we find a $\gamma$-ray line signal at $\sim 43.2$ GeV. In particular, if we take into account only the data in the directions of the Virgo, Fornax, and Ophiuchus clusters, three massive clusters with the highest J-factors, the net TS value is $\approx 30$, corresponding to a global significance of $4.3\sigma$. The signal still presents for the whole sample, though the TS value decreases to $\approx 21$, corresponding to a global significance of $3.7\sigma$, likely because of the lower signal-to-noise ratios for other sources. The absence of this signal in both the inner Galaxy and the Earth limb disfavors an instrumental origin. Anyhow, the canonical dark matter annihilation interpretation is challenged by the stringent constraints set by the inner Galaxy and it is unclear whether there could be a peculiar astrophysical origin. Besides investigating its possible physical origin, it is urgent to clarify the robustness of this signal. The Dark Matter Particle Explorer (DAMPE) also works in this energy range and has an excellent energy resolution. But its acceptance is smaller than Fermi-LAT by a factor of $\sim 7$ \cite{DAMPE:2017cev}. The expected detection prospect of the line signal is low, consistent with the null result found in the analysis of the DAMPE data (see Sec.~\ref{appx:dampe} of the Supplemental Material~\citep{supp}). 
If the 43 GeV $\gamma$-ray line is indeed intrinsic, by 2040, the Fermi-LAT data will be doubled, and the signal will be confirmed. A quicker independent test is also possible. For the VLAST proposed by some DAMPE people,  a robust clarification is expected in the first two years of its performance.  

{\it Acknowledgments---}
This work was supported in part by the National Key Research and Development Program of China (Nos. 2022YFF0503301 and  2022YFF0503304), the Project for Young Scientists in Basic Research of the Chinese Academy of Sciences (Nos. YSBR-092 and YSBR-061), the New Cornerstone Science Foundation through the XPLORER PRIZE, the National Natural Science Foundation of China (Nos. 12220101003, 11921003, U1738210, and 12003074), and the Entrepreneurship and Innovation Program of Jiangsu Province.
This research has made use of data obtained from the High Energy Astrophysics Science Archive Research Center (HEASARC), provided by NASA's Goddard Space Flight Center.
This research has made use of the data resources from the DArk Matter Particle Explorer (DAMPE) satellite mission supported by Strategic Priority Program on Space Science, and China data service provided by the National Space Science Data Center of China.
This research has made use of the CTA instrument response functions provided by the CTA Consortium and Observatory.

\bibliography{refs}

\appendix
\section{The sample of the galaxy clusters}\label{appx:sample}
\begin{table*}
    \centering
    \caption{\label{tab:appx:clusters}
    Parameters of the galaxy clusters.
    The uncertainties of the J-factors come from both the virial radii and the concentrations.
    The ROI radii $\theta_{\rm ROI,0}$ in the baseline analyses are given in ninth column, defined with the central values of $\theta_{200}$.
    The dagger represents an overlapping between the ROIs of Virgo and M49.
    The three clusters below the horizontal line are those contained in the previous works~\citep{Anderson:2015dpc,Liang:2016pvm,Shen:2021fie} but excluded in this work due to the strong background.
    }
    \begin{tabular}{lrrcccccccc}
        \hline\hline
        galaxy cluster & \multicolumn{1}{c}{$\alpha$} & \multicolumn{1}{c}{$\delta$} & $z$ &     $M_{200}$      & $R_{200}$ & $c_{200}$ & $\theta_{\rm 200}$ & $\theta_{\rm ROI,0}$ &  $\log_{10} (J_{\rm NFW})$    & $b_{\rm sh}$  \\
                       & \multicolumn{1}{c}{($\deg$)} & \multicolumn{1}{c}{($\deg$)} &     & ($10^{14}M_\odot$) &($\rm Mpc$)&           &       ($\deg$)     &       ($\deg$)       &   ($\rm GeV^2\,cm^{-5}$)      &               \\
        \hline
    Virgo &$  187.704   $&$  12.391  $&$ 0.0038  $&$ 1.005^{+0.018}_{-0.018}     $&$ 0.926^{+0.005}_{-0.005}     $&$ 8.80^{+0.20}_{-0.20}    $&$ 3.47^{+0.02}_{-0.02}  $&  $3.47^\dagger$ &$ 18.358^{+0.026}_{-0.026}    $&$ 47.56 $ \\
   Fornax &$  54.669    $&$ -35.310  $&$ 0.0046  $&$ 1.196^{+0.522}_{-0.540}     $&$ 0.981^{+0.126}_{-0.178}     $&$ 5.48^{+2.39}_{-1.59}    $&$ 3.02^{+0.38}_{-0.55}  $&  $3.02$          &$ 17.906^{+0.335}_{-0.407}    $&$ 48.53 $ \\
Ophiuchus &$  258.111   $&$ -23.363  $&$ 0.0280  $&$ 34.691^{+22.619}_{-22.310}  $&$ 2.991^{+0.545}_{-0.869}     $&$ 4.98^{+1.94}_{-1.39}    $&$ 1.56^{+0.28}_{-0.46}  $&  $1.56$          &$ 17.769^{+0.404}_{-0.603}    $&$ 67.34 $ \\
      M49 &$  187.444   $&$  7.997   $&$ 0.0038  $&$ 0.441^{+0.020}_{-0.019}     $&$ 0.704^{+0.010}_{-0.010}     $&$ 5.86^{+2.26}_{-1.63}    $&$ 2.62^{+0.04}_{-0.04}  $&  $2.62^\dagger$  &$ 17.685^{+0.258}_{-0.236}    $&$ 43.04 $ \\
    A3526 &$  192.200   $&$ -41.309  $&$ 0.0103  $&$ 3.156^{+0.773}_{-1.329}     $&$ 1.354^{+0.102}_{-0.226}     $&$ 5.20^{+2.18}_{-1.47}    $&$ 1.87^{+0.14}_{-0.31}  $&  $1.87$          &$ 17.600^{+0.275}_{-0.385}    $&$ 54.06 $ \\
    A1060 &$  159.178   $&$ -27.521  $&$ 0.0114  $&$ 2.309^{+0.756}_{-1.011}     $&$ 1.219^{+0.121}_{-0.213}     $&$ 5.28^{+2.25}_{-1.51}    $&$ 1.53^{+0.15}_{-0.27}  $&  $1.53$          &$ 17.388^{+0.302}_{-0.397}    $&$ 52.27 $ \\
     Coma &$  194.947   $&$  27.939  $&$ 0.0232  $&$ 9.003^{+2.585}_{-3.042}     $&$ 1.911^{+0.168}_{-0.245}     $&$ 5.01^{+1.99}_{-1.40}    $&$ 1.19^{+0.11}_{-0.15}  $&  $1.19$          &$ 17.344^{+0.288}_{-0.331}    $&$ 60.05 $ \\
 NGC~4636 &$  190.708   $&$  2.688   $&$ 0.0037  $&$ 0.155^{+0.042}_{-0.062}     $&$ 0.497^{+0.041}_{-0.078}     $&$ 6.35^{+2.79}_{-1.83}    $&$ 1.90^{+0.15}_{-0.30}  $&  $1.90$          &$ 17.313^{+0.297}_{-0.382}    $&$ 37.65 $ \\
     AWM7 &$  43.623    $&$  41.578  $&$ 0.0172  $&$ 4.491^{+1.451}_{-2.167}     $&$ 1.519^{+0.148}_{-0.300}     $&$ 5.12^{+2.16}_{-1.45}    $&$ 1.27^{+0.12}_{-0.25}  $&  $1.27$          &$ 17.308^{+0.299}_{-0.431}    $&$ 56.08 $ \\
    A1367 &$  176.190   $&$  19.703  $&$ 0.0216  $&$ 6.733^{+1.519}_{-2.357}     $&$ 1.736^{+0.122}_{-0.232}     $&$ 5.05^{+2.03}_{-1.41}    $&$ 1.16^{+0.08}_{-0.15}  $&  $1.16$          &$ 17.283^{+0.267}_{-0.338}    $&$ 58.40 $ \\
 NGC~5813 &$  225.299   $&$  1.698   $&$ 0.0064  $&$ 0.385^{+0.464}_{-0.304}     $&$ 0.672^{+0.203}_{-0.273}     $&$ 5.92^{+3.33}_{-1.86}    $&$ 1.49^{+0.45}_{-0.61}  $&  $1.49$          &$ 17.184^{+0.502}_{-0.780}    $&$ 42.32 $ \\
    A2877 &$  17.480    $&$ -45.922  $&$ 0.0241  $&$ 6.166^{+6.902}_{-3.521}     $&$ 1.684^{+0.479}_{-0.414}     $&$ 5.06^{+2.18}_{-1.45}    $&$ 1.01^{+0.29}_{-0.25}  $&  $1.01$          &$ 17.155^{+0.496}_{-0.508}    $&$ 57.90 $ \\
     S636 &$  157.421   $&$ -35.326  $&$ 0.0093  $&$ 0.766^{+0.300}_{-0.136}     $&$ 0.845^{+0.098}_{-0.054}     $&$ 5.64^{+2.25}_{-1.64}    $&$ 1.29^{+0.15}_{-0.08}  $&  $1.29$          &$ 17.126^{+0.324}_{-0.250}    $&$ 46.05 $ \\
     \hline
    A3627 &$  243.555   $&$ -60.843  $&$ 0.0163  $&$ 4.487^{+0.903}_{-1.034}     $&$ 1.519^{+0.096}_{-0.127}     $&$ 5.12^{+2.03}_{-1.43}    $&$ 1.34^{+0.08}_{-0.11}  $&  $\cdots$        &$ 17.353^{+0.260}_{-0.269}    $&$ 56.08 $ \\
  Perseus &$  49.946    $&$  41.515  $&$ 0.0183  $&$ 5.477^{+1.804}_{-2.720}     $&$ 1.622^{+0.162}_{-0.332}     $&$ 5.08^{+2.14}_{-1.43}    $&$ 1.28^{+0.12}_{-0.27}  $&  $\cdots$        &$ 17.337^{+0.301}_{-0.443}    $&$ 57.22 $ \\
    3C129 &$  72.560    $&$  45.026  $&$ 0.0223  $&$ 4.796^{+2.515}_{-2.219}     $&$ 1.550^{+0.234}_{-0.290}     $&$ 5.11^{+2.14}_{-1.46}    $&$ 1.00^{+0.16}_{-0.18}  $&  $\cdots$        &$ 17.117^{+0.357}_{-0.416}    $&$ 56.46 $ \\
        \hline\hline
    \end{tabular}
\end{table*}

In Table \ref{tab:appx:clusters} we summarize the parameters of the target galaxy clusters.
For Virgo we take $R_{200}=974.1\pm5.7~h_{70}^{-1}\,\rm kpc$ \cite{Simionescu2017}. A larger $R_{200}$ has been suggested in recent literature \cite{2020A&A...635A.135K}, but the dark matter beyond such a radius needs to be tiny, which is hard to be understood in the $\Lambda$CDM paradigm.
The concentration parameter $c_{200}=8.80\pm0.20$ in \cite{Simionescu2017} is also adopted.
Ophiuchus is the second brightest galaxy cluster in X-ray band. Because of its low latitude, the optical data are still poor and the inferred density profile in optical is significantly different from that measured in infrared band \cite{2022A&A...663A.158G}, therefore in this work we take the $R_{500}=2.079^{+0.328}_{-0.579}~h_{70}^{-1}\,\rm Mpc$ derived in X-rays \cite{Chen:2007sz}.
The M49, located outside the Virgo cluster, is another good target for dark matter detection due to its proximity.
According to the recent observation, its virial radius is $R_{\rm 200}=740\pm11~h_{70}^{-1}\,\rm kpc$~\citep{2019AJ....158....6S}.
The parameters of the Antlia cluster (S636) are also updated~\citep{Wong2016}, though the angular size $\theta_{\rm 200}$ only slightly changes from $1.27^\circ$~\citep{Chen:2007sz} to $1.29^\circ$.
For the other clusters, we take the redshift $z$ and radius $R_{\rm 500}$ tabulated in \citep{Chen:2007sz}, and then calculate the virial radii $R_{\rm 200}$ and virial masses $M_{\rm 200}$.
The right ascensions $\alpha$ and declinations $\delta$ of the cluster centers in J2000 epoch are taken from \citep{Reiprich:2001zv}.
We use the mass-concentration relation from~\citep{Sanchez-Conde:2013yxa} for the clusters except the Virgo cluster and also consider its $1\sigma$ scatter of 0.14~dex. 
The following cosmological parameters are adopted: $H_0= 73.6~\rm km\,s^{-1}\,Mpc^{-1}$, $\Omega_{\rm M}=0.334$, and $\Omega_\Lambda=0.666$~\citep{Brout:2022vxf}. 

\begin{figure}
    \includegraphics[width=0.48\textwidth]{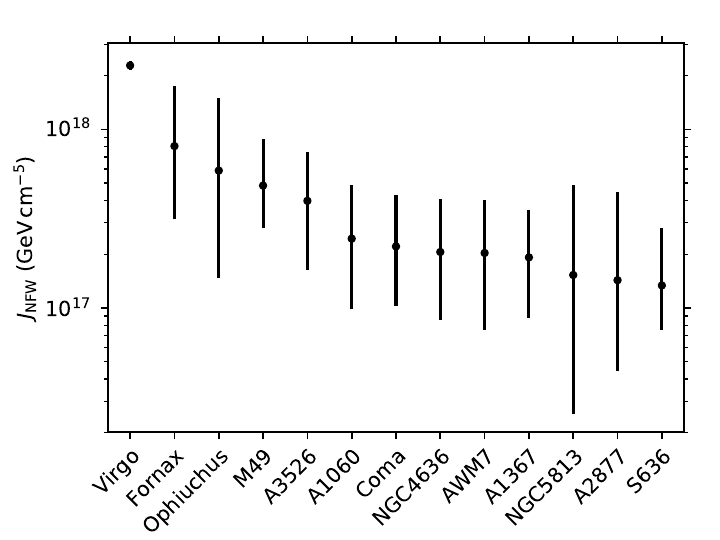}
    \caption{
    The J-factors of the nearby 13 galaxy clusters sorted by the central values.
    The uncertainties of the J-factors stem from the errors of the virial radii and the scattering of the mass-concentration relation.
    }
    \label{fig:appx:jfactors}
\end{figure}

The J-factors within the angular radius $\theta_{\rm 200}$ is
\begin{equation}
    J_{\rm NFW} = \int_0^{\theta_{\rm 200}} 2\pi\theta {\rm d}\theta \int_{\rm los} \rho_{\rm NFW}^2(r(s,\theta))\, {\rm d}s,
    \label{eq::jfactor}
\end{equation}
where $s$ is the length along the line of sight.
$\rho_{\rm NFW}(r)=\rho_0/[(r/r_{\rm s})(1+r/r_{\rm s})^2]$ is the density profile of the cluster.
The scale radius $r_{\rm s}$ can be derived through $r_{\rm s}=R_{\rm 200}/c_{\rm 200}$ and the normalization $\rho_{\rm 0}$ can be achieved using $M_{200}$.
We also propagate the uncertainties of both the virial radii and the concentrations to the J-factors (Fig.~\ref{fig:appx:jfactors}).
The DM density of the galaxy clusters could deviate from the standard NFW profile, as suggested in some observations of brightest cluster galaxies (e.g.~\citep{2013ApJ...765...25N,2025MNRAS.541.2341C}), but the true DM density profiles for the clusters in this work are still unknown.
So we only consider the NFW profile and ignore the uncertainty concerning the density slope in this work.
The last column of the Table \ref{tab:appx:clusters} presents the baseline boost factor parameterized in~\citep{Moline:2016pbm} for the subhalos following the mass function of ${\rm d}n_{\rm sh}/{\rm d}m_{\rm sh} \propto m_{\rm sh}^{-2}$.
Nevertheless, the boost factors can be much larger or smaller due to the systematic uncertainties of the mass-concentration relation and the mass function for the subhalos~\citep{Ishiyama:2019hmh,Ando2019}.

\section{The Unbinned likelihood method}\label{appx:likelihood}
\subsection{Exposure and energy dispersion}\label{appx:likelihood:irf}
We implement the Fermi-LAT instrumental response functions following the documents in the Fermi Science Support Center.\footnote{\url{https://fermi.gsfc.nasa.gov/ssc/data/analysis/lat_irfs/}}
Here, we introduce the parameterization of the exposure and energy dispersion function.
Please note that, in this subsection, to keep consistent with the original Fermi documents, $E$ and $E'$ denote the true and reconstructed energies, different from the other sections.

The exposure $\epsilon(E,\hat{\bm r})$ at the coordinate $\hat{\bm r}$ and true incident energy $E$ for a given event type $s$ is~\citep{Ackermann:2012,Knodlseder2016}
\begin{equation}
\begin{aligned}
    \epsilon(E&,\hat{\bm r}; s)
    = \int \frac{{\rm d}\epsilon}{{\rm d}\theta}(E,\hat{\bm r},\theta; s){\rm d}\theta\\
    &= \int
    \left[p_0(E) \tau_{\rm wgt}(\hat{\bm r},\theta)+p_1(E) \tau(\hat{\bm r},\theta)\right] A_{\rm eff}(E, \theta; s) {\rm d}\theta,
    \label{eq::supp:exposure}
\end{aligned}
\end{equation}
where $A_{\rm eff}(E,\theta; s)$ is the effective area at the energy $E$ and inclination angle $\theta$, which is obtained by interpolating the table {\tt aeff\_P8R3\_ULTRACLEAN\_V3\_EDISP.fits}.
To account for reduced efficiency in the event reconstruction caused by the ghost events, the efficiency factors $(p_0, p_1)$ are introduced.
They are parameterized with piecewise linear functions as a function of $\lg(E/{\rm MeV})$.\footnote{\url{https://fermi.gsfc.nasa.gov/ssc/data/analysis/lat_irfs/irf_effective_area.html}}
$\tau(\hat{\bm r},\theta)$ and $\tau_{\rm wgt}(\hat{\bm r},\theta)$ in Eq.(\ref{eq::supp:exposure}) are the total live time and the exposure-weighted live time on the sky $\hat{\bm r}$ at a given inclination angle $\theta$, respectively, which can be calculated with {\tt gtltcube}.
If there are multiple event types, the total exposure is the sum of the individual exposures: $\epsilon(E,\hat{\bm r}) = \sum_s\epsilon(E,\hat{\bm r}; s)$.
We compare our all-sky exposure map from 100~MeV to 1~TeV with that from {\tt gtexpcube2} and find that the two maps are well consistent.

The energy dispersion function is parameterized with\footnote{\url{https://fermi.gsfc.nasa.gov/ssc/data/analysis/lat_irfs/irf_energy_dispersion.html}}
\begin{equation}
\begin{aligned}
    D(E'; E,\theta, s) &= F\times g(x_E; \sigma_1,k_1, b_1, p_1) \\
             &+ (1-F)\times g(x_E; \sigma_2,k_2, b_2, p_2),
    \label{eq::supp:edisp}
\end{aligned}
\end{equation}
where $x_{E}(E,E',\theta; s)=(E'-E)/(E\cdot S_{D}(E,\theta; s))$ is the scaled energy difference, and the scaling factor is
\begin{equation}
\begin{aligned}
        S_D(E,\theta; s) &= c_0 \lg^2 (E/{\rm MeV}) + c_1 \cos^2\theta + c_2\lg (E/{\rm MeV}) \\
                &+ c_3\cos\theta + c_4 \lg (E/{\rm MeV}) \cdot \cos\theta + c_5,\nonumber
\end{aligned}
\end{equation}
where the parameters $(c_0, c_1, c_2, c_3, c_4, c_5)$ for event type $s$ are given in {\tt edisp\_P8R3\_ULTRACLEAN\_V3\_EDISP.fits}.
$g(x_E)$ is the asymmetric exponential power function
\begin{equation}
\begin{aligned}
    g(x_E&; \sigma_i,k_i, b_i, p_i) = \frac{p_i}{\sigma_i \Gamma(1/p_i)}\frac{k_i}{1+k_i^2} \\
            \times&
    \begin{cases}
         \exp\left[-(\frac{k_i}{\sigma_i}|x_E-b_i|)^{p_i} \right] & \quad x_E-b_i \geq 0,\\
         \exp\left[-(\frac{1}{k_i\sigma_i}|x_E-b_i|)^{p_i}\right] & \quad x_E-b_i < 0,\nonumber
    \end{cases}
    \end{aligned}
\end{equation}
where $\Gamma(\cdot)$ is the Gamma function.
The values of the parameters $(F, \sigma_1,k_1, b_1, p_1, \sigma_2,k_2, b_2, p_2)$ can be obtained by interpolating the table in the {\tt fits} file.
Finally, the energy dispersion function given the true energy $E$ at the sky position $\hat{\bm r}$ for multiple event types is
\begin{equation}\label{eq::appx:edisp_func_wt}
    D(E'; E, {\hat{\bm r}})= \frac{\sum_s \int\frac{{\rm d}\epsilon}{{\rm d}\theta}(E,\hat{\bm r},\theta; s)D(E'; E,\theta, s){\rm d}\theta}{\sum_s \epsilon(E,{\hat{\bm r}}; s)},
\end{equation}
with $s$ representing different event types.
Our implementation can well reproduce the acceptance weighted energy resolution provided by Fermi-LAT.\footnote{\url{https://www.slac.stanford.edu/exp/glast/groups/canda/lat_Performance.htm}}

\subsection{The model and the likelihood function}
In this work, we collect all the photons within the ROIs of the clusters and search for the spectral features in this merged dataset.
The unbinned likelihood method is adopted, and the likelihood function is defined as~\citep{Cash1979}
\begin{equation}
    \ln \mathcal{L}(\Theta) = \sum_{i=1}^{N_{\rm ph}} \ln (\lambda(E_i;\Theta)) - \int_{E_{\rm  min}}^{E_{\rm max}} \lambda(E;\Theta) {\rm d}E,
\end{equation}
where $\lambda(E; \Theta)$ is the total expected photon count spectrum of the merged data set given the spectral parameters $\Theta$.
The expected number of counts is $n^{\rm pred}= \int \lambda(E) {\rm d}E$.
$N_{\rm ph}$ is the number of all the photons within the ROIs of the clusters in the energy range from $E_{\rm min}$ to $E_{\rm max}$.

For the merged data set, the expected count spectrum for the $j$-th component at energy $E$ would be
\begin{equation}\label{eq::appx:total_model}
\begin{aligned}
    \lambda_j (E&; \Theta_j) 
    = \sum_{i=1}^{N_{\rm gcls}} \int_{{\rm ROI}_i} \mathcal{M}_{j}(E, \hat {\bm r}) \epsilon(E,\hat {\bm r}) {\rm d}\Omega\\
    &\approx \sum_{i=1}^{N_{\rm gcls}} S_{ij}(E) \tilde\epsilon_i(E) \Omega_i 
    = S_j(E; \Theta_j) \sum_{i=1}^{N_{\rm gcls}} (\tilde\epsilon_i(E) \Omega_i) \\
    &= S_j(E; \Theta_j) \Omega_{\rm tot} \bar{\epsilon}(E)
    = F_j(E; \Theta_j)\bar{\epsilon}(E), 
\end{aligned}
\end{equation}
where $\mathcal{M}_{j}(E, \hat {\bm r})$ is the $\gamma$-ray radiation model of the $j$-th component,
$S_{ij}(E
)$ is the spectrum of the $j$-th component within the $i$-th ROI,
$\Omega_i$ is the solid angle for the $i$-th cluster,
$\tilde\epsilon_i(E)=\int_{{\rm ROI}_i} \epsilon_i(E, \hat{\bm r}) {\rm d}\Omega/\Omega_i$ is the ROI-averaged exposure,\footnote{
$\tilde\epsilon_i(E)$ is a good approximation for the small ROIs of the clusters due to the rather uniform exposure of the Fermi-LAT as well as the small size of sources.
Our tests show that the count spectrum is changed by $\sim 0.1\%$ using this approximation for the Galactic diffuse emission model within a ROI, which does not affect the TS values of the line.
}
$\Omega_{\rm tot}=\sum_i \Omega_i$, $S_j(E; \Theta_j)$ is the weighted spectrum of all the clusters with parameters $\Theta_j$, 
$\bar{\epsilon}(E)=\sum_i (\tilde\epsilon_i(E) \Omega_i)/\sum_i  \Omega_i$, and $F_j(E; \Theta_j)=S_j(E; \Theta_j) \Omega_{\rm tot}$. 
For the line component, the spectrum is~\citep{Liang:2016pvm,Shen:2021fie}\footnote{We have also tested with additional J-factor weights
and find that the TS values of the line do not change.
Please see Sec.~\ref{appx:containment}.
}
\begin{equation}
\begin{aligned}
        F_{\rm s}(E)
        &\propto \bar{D}(E; E_{\rm line})\\
        &= \frac{\sum_i \int_{{\rm ROI},i}\epsilon(E_{\rm line},{\hat{\bm r}})D(E; E_{\rm line}, {\hat{\bm r}}){\rm d}\Omega}{\sum_i \int_{{\rm ROI},i} \epsilon(E_{\rm line}, {\hat{\bm r}}) {\rm d}\Omega },
\end{aligned}
\end{equation}
where $D(E; E', {\hat{\bm r}})$ is the energy dispersion function given the true energy $E'$ at the sky position ${\hat{\bm r}}$ (see Sec.~\ref{appx:likelihood:irf}).
Its count spectrum is $\lambda_{\rm s}(E)=F_{\rm s}(E)\bar{\epsilon}(E_{\rm line})$.
For the background component, the total spectrum is assumed to follow the power-law spectrum within a narrow energy window, $F_{\rm b}(E) \propto E^\Gamma$ and $\lambda_{\rm b}(E)=F_{\rm b}(E)\bar{\epsilon}(E)$.
The total count spectrum is the sum of the components
$\lambda(E;\Theta) = \sum_j \lambda_j(E; \Theta_j)$.
Please note that our treatments on $F_{\rm s}(E)$, $F_{\rm b}(E)$, and the exposure are the same as that adopted by the Fermi-LAT collaboration in~\citep{Fermi-LAT:2015kyq}.

\section{Global significance of the line}\label{appx:trials}
\begin{figure}
	\centering  
     \includegraphics[width=0.48\textwidth]{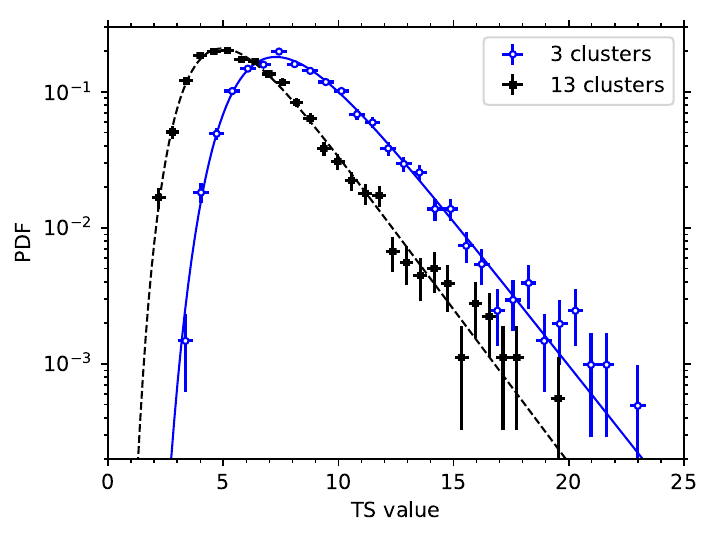}
    \caption{
    The probability distributions of the maximal TS values of 3000 random blank sky simulations for the whole sample (the black points) and the first three galaxy clusters (the blue points).
    The $y$ axis values show probability density, proportional to the number of simulations $N_i$ with the maximum TS values in the $i$-th bin, and the error bars are the Poisson uncertainties.
    The lines are the best-fit trial-corrected $\chi^2$ distribution obtained via a Poisson likelihood fit to the histogram of $N_i$.
    }
\label{fig:appx:trials}
\end{figure}

\newcommand{\ls}{$^\dagger$} 
\newcommand{\gs}{$^\star$}   
\begin{table*}
  \centering
  \small
  \caption{\label{tab::appx:evclasstype3}
      The peak TS value of the linelike signal around 43~GeV for the first three clusters.
      $N_{\rm obs}$ is the observed counts between 40~GeV and 46~GeV.
      The superscript at TS value indicates the best line energy deviates from $43.2~\rm GeV$ in the sliding windows analyses:
      the dagger and the star represent the $\gamma$-ray lines at $42.2~\rm GeV$ and $44.3~\rm GeV$ respectively.
  }
  \begin{tabular}{l|cc|cc|cc|cc|cc}
   \hline\hline
    \multirow{2}{*}{Event Class} & \multicolumn{2}{c|}{EDISP(0+1+2+3)} & \multicolumn{2}{c|}{EDISP(1+2+3)} & \multicolumn{2}{c|}{EDISP(2+3)} & \multicolumn{2}{c|}{FRONT} & \multicolumn{2}{c}{BACK} \\
    \cline{2-11}
    \multirow{2}{*}{} & TS & $N_{\rm obs}$ & TS & $N_{\rm obs}$ & TS & $N_{\rm obs}$ & TS & $N_{\rm obs}$ & TS & $N_{\rm obs}$ \\
    \hline
    SOURCE         &  18.3     &  58  &  25.2     &  51  &  16.3     &  37  &  14.8\gs & 39 &  4.1    & 19 \\
    SOURCEVETO     &  17.6\gs  &  53  &  26.7     &  47  &  16.8     &  33  &  14.9\gs & 34 &  4.5    & 19 \\
    CLEAN          &  18.0     &  51  &  26.0     &  45  &  15.8     &  33  &  13.0\gs & 32 &  5.6    & 19 \\
    ULTRACLEAN     &  22.5     &  50  &  30.1     &  44  &  18.6     &  32  &  13.4    & 31 &  9.2\ls & 19 \\
    ULTRACLEANVETO &  17.9     &  44  &  23.6     &  39  &  13.7     &  28  &  10.6\gs & 26 &  7.5\ls & 18 \\
    \hline\hline
  \end{tabular}
\end{table*}

\begin{table*}
  \centering
  \small
  \caption{\label{tab::appx:evclasstype13}
      The TS value of the linelike signal around 43~GeV for the 13 clusters.
  }
  \begin{tabular}{l|cc|cc|cc|cc|cc}
   \hline\hline
   \multirow{2}{*}{Event Class} & \multicolumn{2}{c|}{EDISP(0+1+2+3)} & \multicolumn{2}{c|}{EDISP(1+2+3)} & \multicolumn{2}{c|}{EDISP(2+3)} & \multicolumn{2}{c|}{FRONT} & \multicolumn{2}{c}{BACK} \\
  \cline{2-11}
   \multirow{2}{*}{} & TS & $N_{\rm obs}$ & TS & $N_{\rm obs}$ & TS & $N_{\rm obs}$ & TS & $N_{\rm obs}$ & TS & $N_{\rm obs}$ \\
    \hline
    SOURCE         &  16.0\gs  &  90  &  19.4     &  79  &  11.9     &  55  & 12.5\gs & 58 &  4.1\gs & 32 \\
    SOURCEVETO     &  15.4\gs  &  79  &  21.2     &  71  &  13.2     &  49  & 13.1\gs & 50 &  3.8\ls & 29 \\
    CLEAN          &  16.5\gs  &  80  &  19.5     &  70  &  12.2     &  49  & 10.7\gs & 48 &  5.8\gs & 32 \\
    ULTRACLEAN     &  17.5\gs  &  73  &  21.3     &  64  &  13.7     &  45  &  9.0\gs & 42 &  8.9\gs & 31 \\
    ULTRACLEANVETO &  14.8\gs  &  64  &  17.0     &  56  &  10.9\ls  &  38  &  8.0\gs & 35 &  7.2\ls & 29 \\
    \hline\hline
  \end{tabular}
\end{table*}

Since multiple trials are adopted to search for the largest TS values, the ``look elsewhere effect'' should be considered.
To estimate the global significance, we perform the random sky simulations~\citep{Conrad2015}.
In each simulation, 13 regions are selected randomly in the sky with the radii aligned with the galaxy clusters.
In order to ensure the properties of the random regions not too different from the source regions,
we exclude those around the Galactic plane ($|b|<5^\circ$) or the Galactic center ($|l|<20^\circ$ and $|b|<9^\circ$), or close to the strong point sources ($F_{1000}>2\times 10^{-9}~\rm ph\,cm^{-2}\,s^{-1}$ in the 4FGL-DR4 catalog~\citep{Fermi-LAT:2022byn}).
In total, 3000 random ROI sets are created.
Considering the photons can be reused multiple times in different simulations, we randomize the energies of the events in the ROIs according to the energy dispersion function at the source.\footnote{We perform a test using the simulations without randomizing the energies and find that the global significance barely changes.
We also verify the global significance of the 13 clusters using 3000 MC simulations based on the best-fit null model assuming a LogParabola background spectrum (see Sec.~\ref{appx:systematics:bkgtype}), the global significance is $3.7\sigma$, consistent with the baseline result.}
In this way, even if the same photon is presented in several simulations, the energies are different.

For the whole sample, 
the local TS value is 21.3 according to the sliding-window technique.
We should take the trials over the line energies into account.
We carry out the same sliding windows analyses using all the 13 random regions in the each simulation and record the largest TS value.
The probability distribution of the TS value is shown with the black points in Figure~\ref{fig:appx:trials}.
The points are $(N_i/\sum_j N_{j}) / \Delta {\rm TS}_i$, where $N_i$ is the number of simulations whose maximum TS values falls in the $i$-th bin, $\Delta {\rm TS}_i$ is the bin width, and $\sum_j N_{j}=3000$ is the total number of simulations.
The vertical error bars are $(\sqrt{N_i}/\sum_j N_{j}) / \Delta {\rm TS}_i$, considering $N_i$ follows the Poisson distribution.
We conduct a Poisson likelihood fit to the histogram $N_i$ with the model following the density profile of
the trial-corrected $\chi^2$ distribution, whose cumulative distribution function 
$P_{\rm g}(Q<c)$ for the statistic $Q$ is assumed to be~\citep{Weniger:2012tx}
\begin{equation}
    P_{\rm g}(Q<c; k,t) = \left[ P(\chi^2_k<c) \right]^t.
    \label{eq::appx:trial}
\end{equation}
The best-fit trial factor is $t=27 \pm 6$ and the degree of freedom is $k=1.27\pm0.16$.
The global significance is $\approx 3.7\sigma$ for ${\rm TS}=21.3$.
We further cross-check the global significance using the upcrossing method~\citep{Gross2010}.
This method can set the lower bound of the global significance by counting the number of times the TS value curve crosses the reference level ($c_0$) upward, $N(c_0)$.
The reference level is set to be $c_0=0.5$ according to~\citep{Gross2010}.
We count the number of upcrossings in the same blank sky simulation samples, which gives $\left< N(c_0)\right>=10.1\pm1.7$.
Then the global $p$-value can be bounded by~\citep{Gross2010}
\begin{equation}
P_{\rm g}(Q>c) \leq P(\chi^2_1>c) + \left< N(c_0) \right> \exp\left(-\frac{c-c_0}{2}\right).
\end{equation}
Therefore the global significance is $\geq 3.6\sigma$, consistent with the $3.7\sigma$ global significance from the trial-corrected $\chi^2$ distribution.

For the first three galaxy clusters, there is another trial over the number of analyzed sources besides the line energies.
In each simulation, we derive the accumulative TS values for gradually increasing samples whose ROIs follow those of target clusters.
The largest TS value among all the fittings in each simulation is collected and the final probability distribution is shown in blue in Figure~\ref{fig:appx:trials}.
We fit the distribution with the trial-corrected $\chi^2$ distribution defined in Eq.(\ref{eq::appx:trial}).
The best-fit parameters are $t=28 \pm 4$ and $k=2.32\pm0.17$.
Therefore, the TS value of 30.1 corresponds to a global significance of $\approx 4.3\sigma$.

\section{Systematic uncertainties}\label{appx:systematics}
\subsection{Event classes and event types}\label{appx:systematics:evclass_type}

\begin{figure*}
    \includegraphics[width=0.48\textwidth]{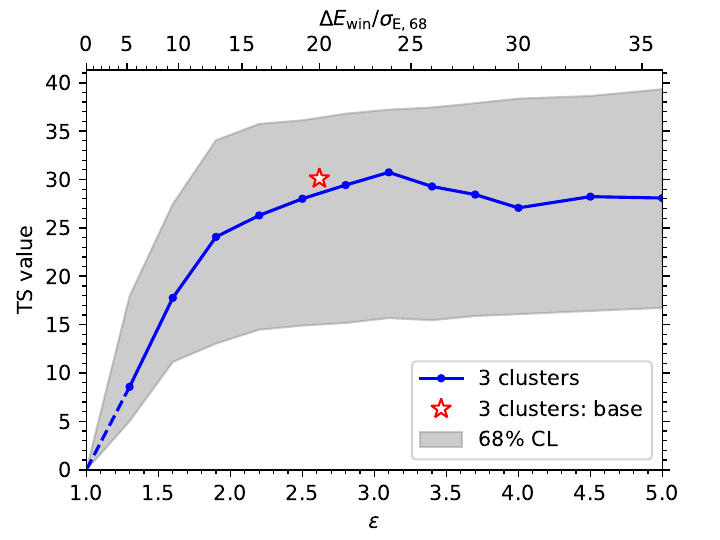}
    \includegraphics[width=0.48\textwidth]{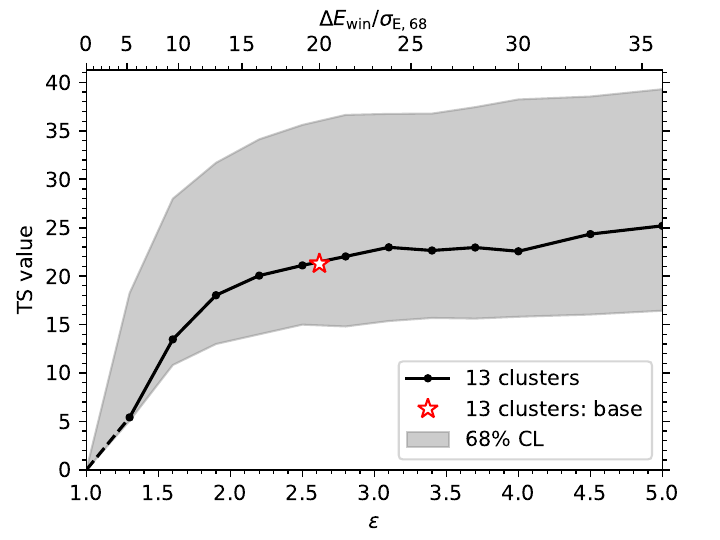}
    \caption{
    The TS values for the energy windows centering at $43.2~\rm GeV$ with various widths.
    The solid lines in the left and right panels are TS values obtained from the top three clusters and the whole sample, respectively.
    The red star is the TS value in the baseline window locating at the equivalent parameter $\varepsilon_{\rm b}$.
    The gray shaded area is the 68\% C.L. band obtained from the MC simulation.
    The parameter $\varepsilon$ in the lower axis defines the window size (Eq.(\ref{eq:appx:win_bound})), while the upper axis demonstrates the size in unit of energy resolution.
    }
    \label{fig:appx:winsize}
\end{figure*}

The Fermi-LAT data are classified into several event classes to balance the effective area and the cosmic-ray background rate.
For each event class, the data are further separated into four event types (EDISP0-3) according to the energy reconstruction quality.
To test the reality of the line signal, we carry out the same unbinned sliding windows analyses for the data set with various event classes and event types.

The TS values of the $\gamma$-ray line around 43~GeV for the three (13) clusters are shown in Table~\ref{tab::appx:evclasstype3} (Table~\ref{tab::appx:evclasstype13}).
No matter what event class, the EDISP(1+2+3) event type data have the largest TS values.
It is reasonable since this event type has the best sensitivity:\footnote{The sensitivity of the line can be estimated with the acceptance divided by the energy resolution~\citep{Xu2022}.}
Removing the EDISP0 data improves the exposure-weighted energy resolution from $\sim 6.3\%$ to $\sim 4.6\%$ ($\sim 36\%$ improvement) at the cost of $\sim 25\%$ effective area.
However, the energy resolution improvement can not compensate for the effective area loss when more event types are dropped.
The front-converting data have larger TS values than the back-converting data because the former has a larger exposure and better energy solution than the latter.
On the other hand, the event class only mildly affects the TS values due to the similar line sensitivity.
Among these event classes, the TS values for the ULTRACLEANVETO data are the lowest.
The data set slightly reduces cosmic-ray background ($\sim 1.5\%$ of the isotropic component at 43~GeV) at the cost of $\sim 13\%$ effective area compared with the ULTRACLEAN data set.

In general, the 43~GeV line exists in various event classes and event types.
For the data sets with the best line sensitivity, which have the event class of ULTRACLEAN, CLEAN, and SOURCEVETO and the event type of EDISP(1+2+3), the TS values of the 43~GeV line are $\sim 27$ and $\sim 21$ for the three clusters and 13 clusters, respectively.

\subsection{Sliding window size}\label{appx:systematics:winsize}

The energy window size is a crucial super parameter in the sliding window analysis.
A narrow window can not well constrain the background, while a large window is more likely to suffer from the systematic uncertainty from the background spectral model.
To illustrate the impact of the window size, we test a series of energy boundaries $[E_{\rm l}, E_{\rm u}]$, where
\begin{equation}
    E_{\rm l}=E_{\rm line}/\sqrt{\varepsilon} {\rm ~and~} E_{\rm u}=E_{\rm line} \sqrt{\varepsilon}.
    \label{eq:appx:win_bound}
\end{equation}
The parameter $\varepsilon$ quantifies the width of the energy window~\citep{Weniger:2012tx}.
We focus on the energy window centering around 43.2~GeV and vary $\varepsilon$ from 1.3 to 5, which can cover the window size from $\sim 5$ times to $\sim 35$ times the 68\% containment of the energy dispersion $\sigma_{E,68}$ ($\approx 2~\rm GeV$ at 43~GeV).
The background component is still assumed to be power-law within the window.

The TS values of the 43.2~GeV line are shown with the solid line in Fig.~\ref{fig:appx:winsize}.
In both cluster samples, the TS values increase when the window is narrow and then stabilize to 28 (three clusters) and 25 (13 clusters) when $\varepsilon > 2$.
To understand the statistical behavior for different window widths, we make 300 MC simulations based on the best-fit alternative spectral models from the energy window with size $\varepsilon=6$, derive the TS values for various window sizes, and calculate their 68\% CL regions as shown in the gray band.
The observed TS value well follows results from the simulations.
We mark the baseline results with the red star in the figures.
Even though the bounds of the baseline energy window are different from those defined with Eq.(\ref{eq:appx:win_bound}),\footnote{
The baseline window does not follow the definition in Eq.(\ref{eq:appx:win_bound}).
We draw the TS value of the baseline window at the equivalent $\varepsilon_{\rm b}$ ($\approx 2.618$).
Such equivalent window shares the same width as the baseline one, which ensures the background counts similar to those using the baseline energy window and thereby leads to a similar TS value.}
the TS values are consistent with those in the plateau, proving the robustness of the baseline results against the energy window definition.

\subsection{Background spectral model}\label{appx:systematics:bkgtype}
\begin{table}
  \centering
  \small
  \caption{\label{tab::appx:bkgtypes}
      The maximum likelihood value $\mathcal{L}$ using the data from 2.5~GeV to 450~GeV for various spectral models.
      $n_{\rm dof}$ is the number of free parameters.
  }
  \begin{tabular}{l|c|c|c}
  \hline\hline
  \multirow{2}{*}{Spectral Model} & \multirow{2}{*}{$n_{\rm dof}$} & \multicolumn{2}{c}{$-2\ln \mathcal{L}$} \\
  \cline{3-4}
  \multirow{2}{*}{} & \multirow{2}{*}{} & 3 clusters  & 13 clusters \\
    \hline
    PL             & 2 &  $17456.3$  &  $16952.0$    \\
    PLEC           & 3 &  $17454.6$  &  $16948.0$   \\
    LogP           & 3 &  $17454.5$  &  $16949.6$    \\
    BPL            & 4 &  $17452.1$  &  $16944.9$   \\
    \hline
    PL + Line      & 4 &  $17434.2$  &  $16929.1$ \\
    PLEC + Line    & 5 &  $17432.2$  &  $16924.2$ \\
    LogP + Line    & 5 &  $17430.1$  &  $16923.5$    \\
   BPL + Line      & 6 &  $17428.5$  &  $16923.5$   \\
    \hline\hline
  \end{tabular}
\end{table}
\begin{table*}[!t]
  \centering
  \small
  \caption{\label{tab::appx:bkgtypes_dip}
      The best-fit models with a dip using the $2.5- 450~\rm GeV$ data from the first three clusters.
      $n_{\rm dof}$ is the number of free parameters.
  }
  \begin{tabular}{l|c|c|c|c|c|c}
  \hline\hline
  Spectral Model & $n_{\rm dof}$ & $-2\ln \mathcal{L}$ & $\hat{E}_{\rm line}$ & $N_{\rm abs}$ & $E_{\rm abs}$ & $\sigma_{\rm abs}$ \\
                 &               &                     &    (GeV)       &               &    (GeV)      &  (GeV) \\
    \hline
    PL + Dip             & 5 &  $17444.7$   &  $\cdots$ & $1.8\pm1.0$ & $63\pm4$ & $6.1\pm1.9$ \\
    LogP + Dip           & 6 &  $17444.5$   &  $\cdots$ & $1.7\pm1.0$ & $63\pm4$ & $5.9\pm1.9$ \\
    BPL + Dip            & 7 &  $17442.3$   &  $\cdots$ & $1.7\pm1.0$ & $62\pm4$ & $5.8\pm1.9$ \\
    \hline
    PL + Dip + Line      & 7 &  $17423.6$   &  $43.3$ & $1.2\pm0.8$ & $62\pm5$ & $9\pm6$ \\
    LogP + Dip + Line    & 8 &  $17422.3$   &  $43.3$ & $1.2\pm0.8$ & $62\pm4$ & $8\pm5$ \\
    BPL + Dip + Line     & 9 &  $17422.1$   &  $43.4$ & $1.1\pm0.7$ & $61\pm4$ & $8\pm4$ \\
    \hline\hline
  \end{tabular}
\end{table*}

\begin{figure}
    \includegraphics[width=0.48\textwidth]{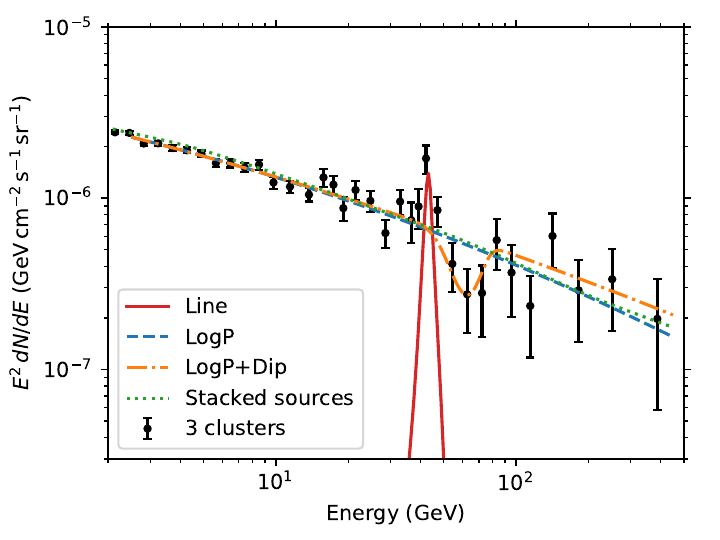}
    \caption{
        The spectrum of the first three galaxy clusters.
        The black points are the observed SED.
        The red solid line shows the $\sim 43$~GeV line model.
        The blue dashed and orange dot-dashed lines represent the optimal LogParabola model and LogParabola+Absorption model, respectively.
        The green dotted line gives the stacked spectrum from the Fermi-LAT $\gamma$-ray sources within the ROIs.
    }
    \label{fig:appx:flux_3src_gl}
\end{figure}

We extend the analyses to the energy range from 2.5~GeV to 450~GeV.
In such a broad-band analysis, the background spectral model is an important component of the systematic uncertainty.
The following spectral types are adopted to substitute the PowerLaw (PL) background: PowerLawExpCutoff (PLEC), LogParabola (LogP), and BrokenPowerLaw (BPL).\footnote{The definitions can be found in \url{https://fermi.gsfc.nasa.gov/ssc/data/analysis/scitools/source_models.html}.}
As displayed in Table~\ref{tab::appx:bkgtypes}, the alternative background spectra are only mildly better than the PowerLaw model.
The excess at $\sim 43~\rm GeV$ can not be explained with the edge of the BrokenPowerLaw model~\citep{Profumo2012}, considering $-2\Delta \ln \mathcal{L}$ just 4.2 and 7.1 for the three clusters and 13 clusters.
On the other hand, adding a line component greatly improves the fitting.
According to the Akaike information criterion ${\rm AIC}\equiv 2n_{\rm dof}-2\ln \mathcal{L}$~\citep{Akaike1974}, the model with the LogParabola background and the $\gamma$-ray line, as shown in Fig.~\ref{fig:appx:flux_3src_gl}, is the best for both the cluster samples.
Furthermore, after comparing the best-fit LogParabola model for the three clusters (blue dashed line) with the stacked spectrum from the Fermi-LAT $\gamma$-ray sources~\citep{Fermi2023_4FGLDR4,Fermi2016_GDE} within the ROIs (green dotted line in Fig.~\ref{fig:appx:flux_3src_gl}), we find the LogParabola spectral model well consistent with the stacked spectrum.
Given the LogParabola background, the TS values of the $\sim 43$~GeV line are 24.4 and 26.1 for the three clusters and 13 clusters, respectively.
They are consistent with the results derived by varying the window size as illustrated in Fig.~\ref{fig:appx:winsize}.

Noticing a weak dip in the spectrum of the first three clusters in Fig.~\ref{fig:appx:flux_3src_gl}, we further quantify the influence of the dip structure on the line signal.
We multiply the previous background spectral model by the following absorption term
\begin{equation}
    M(E) = \left\{ 1 + N_{\rm abs} \exp \left[ -\frac{(E-E_{\rm abs})^2}{2\sigma_{\rm abs}^2} \right]\right\}^{-1},
\end{equation}
where $N_{\rm abs}$, $\sigma_{\rm abs}$ and $E_{\rm abs}$ are the strength, width, and energy center of the dip, respectively.
The fitting results are presented in Table~\ref{tab::appx:bkgtypes_dip}.
The model LogP$+$Dip$+$Line performs best among the alternative models according to the AIC, which is shown with the orange dot-dashed line in Fig.~\ref{fig:appx:flux_3src_gl}.
The dip is located at $\sim 62~\rm GeV$ with a width of $\sim 8$~GeV, wider than the energy resolution at 62~GeV ($\sigma_{E,68}\approx 3~\rm GeV$).
The structure is not significant at present given ${\rm TS_{dip}}\equiv -2\ln (\mathcal{L}_{\rm bkg}/\mathcal{L}_{\rm bkg+dip})=7.8$ if compared to the null hypothesis of LogParabola$+$Line, which corresponds to a local significance of $2.0\sigma$ for three degrees of freedom~\citep{Wilks:1938dza}.
Due to its low significance, the dip may be a statistical fluctuation.
Nevertheless, the TS value of the 43~GeV line is still 22.2 if the LogParabola$+$Dip model is assumed as the background, slightly smaller than the model without the dip but still rather significant.

\subsection{Artifacts in specific incidence angles}\label{appx:systematics:thetaphi}

\begin{figure}
    \includegraphics[width=0.45\textwidth]{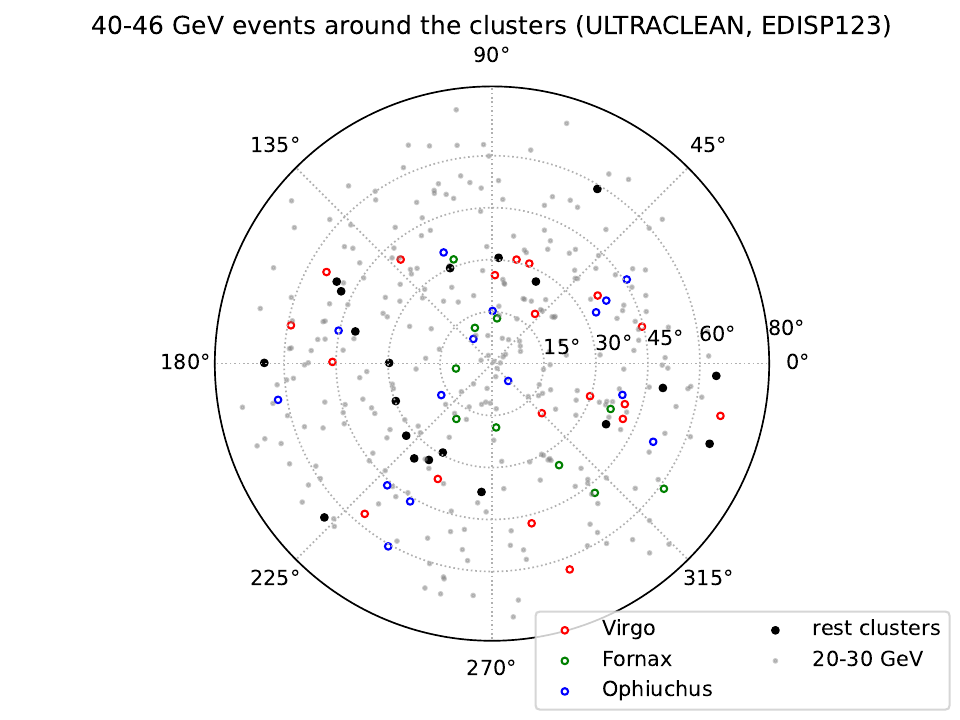}
    \caption{
    The distribution of the $40-46$~GeV photons within the virial radii of the clusters in the instrumental coordinate (inclination angle $\theta$ and azimuthal angle $\phi$).
    The red, green, blue, and black points represent the events from the Virgo, Fornax, Ophiuchus, and the rest 10 clusters, respectively.
    The light gray dots show the distribution of the low-energy events from the clusters.
    }
    \label{fig:appx:thetaphi_ev}
\end{figure}

\begin{figure}
    \includegraphics[width=0.45\textwidth]{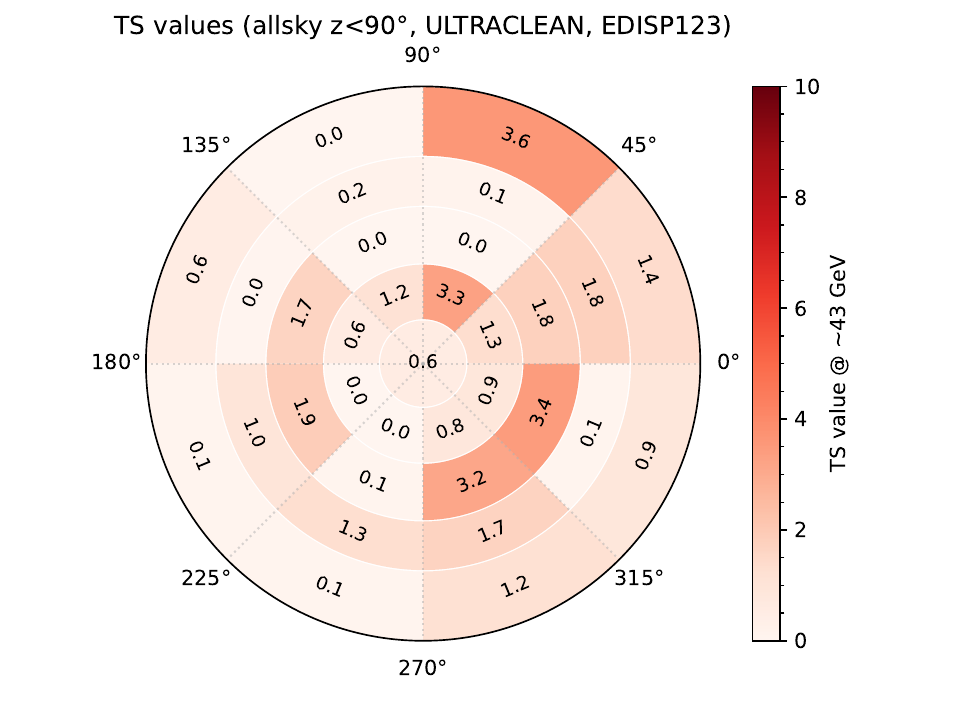}
    \caption{
    The TS value of the $\sim$43~GeV line in various incident angle bins of the all-sky ULTRACLEAN EDISP(1+2+3) data.
    The edges of the $\theta$ bins are $0.0^\circ$, $12.9^\circ$, $29.0^\circ$, $45.6^\circ$, $60.0^\circ$ and $78.5^\circ$.
    The numbers are the TS values of the $\gamma$-ray line in the ($\theta,\phi$) bins.
    }
    \label{fig:appx:thetaphi_ts}
\end{figure}

The Ophiuchus and NGC~4636 are close to the ecliptic plane (the ecliptic latitude $|\beta|<10^\circ$).
Since the telescope keeps the solar panels aligned to the Sun, the incident angles in instrumental coordinate (inclination angle $\theta$ and azimuthal angle $\phi$) for these two sources are restricted in a narrow range around $\phi\sim 0^\circ$ or $\phi\sim 180^\circ$~\citep{Finkbeiner2013}, which might produce artifacts in energy spectrum due to the problematic instrumental responses in the specific angles.
In Fig.~\ref{fig:appx:thetaphi_ev}, we present the incident angles of the $40-46$ GeV events from the galaxy clusters in the instrumental coordinate.
The 44 (64) photons from the first three clusters (all 13 clusters) can well follow the distribution in the lower energy range.
No obvious pattern is present in the instrumental coordinate, therefore the 43~GeV line candidate may not originate from the instrumental artifact in specific incident angles.

To further investigate the systematic uncertainty about the incident angle, we choose the all-sky ULTRACLEAN EDISP(1+2+3) data with zenith angles less than $90^\circ$, partition the data into $(\theta,\phi)$ bins, and do the same analyses for the windows centering at $\sim 43~\rm GeV$.
We consider the instrumental response functions (IRFs) for the corresponding inclination angle in the analyses.
The TS values of the 43~GeV line in various incident angles are demonstrated in Fig.~\ref{fig:appx:thetaphi_ts}.
The largest TS value is merely 3.6, suggesting no significant structure located at $\sim 43~\rm GeV$.
Furthermore, there are at least 2600 events in the $(\theta,\phi)$ bins, which are 10 (6) times the observed counts of the three clusters (the 13 clusters).
If the 43~GeV line in the galaxy cluster comes from the artifacts of the effective area, the TS values should be much larger than those in Fig.~\ref{fig:appx:thetaphi_ts}.
Therefore, the 43~GeV line may not be result from the artifacts in specific incident angles.

\subsection{The energy-dependent variation of effective area}\label{appx:systematics:fsys}
\begin{figure}
    \includegraphics[width=0.48\textwidth]{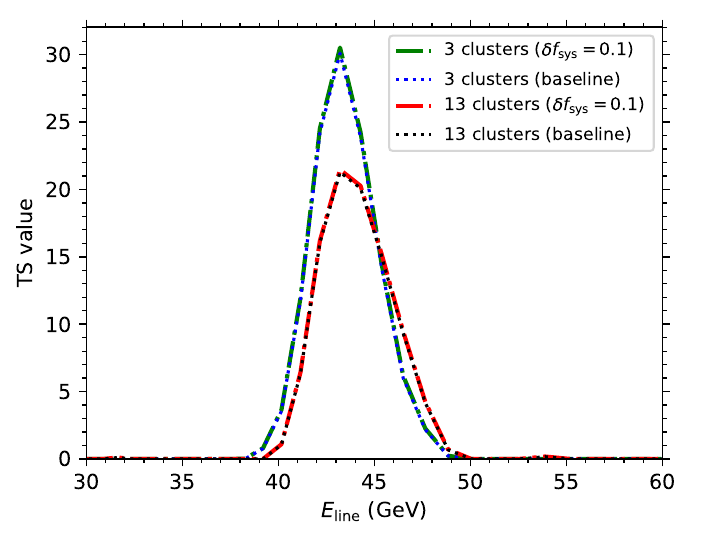}
    \caption{
    The TS values of the line-like excess considering the variation of the effective area (dot-dashed lines).
    The green (red) dot-dashed line shows the result of the three clusters (all 13 clusters) considering the fractional amplitude of the variation $\delta f_{\rm sys}=10\%$.
    We also present the baseline results of the two samples with the dotted lines.
    }
    \label{fig:appx:fsys_ts}
\end{figure}

The effective area in the {\tt Fermitools} is partitioned into 16 energy bins per decade.
Unmodeled energy-dependent variation of the effective area may exist within a single energy bin and is likely to create a spurious sharp structure in the spectrum.
According to the analyses of the Earth limb, Vela pulsar, and the Galactic plane~\citep{Fermi-LAT:2015kyq,Cheng:2023chi}, the fractional amplitude of the variation $\delta f_{\rm sys}$ is less than 2\% from 100~MeV to 100~GeV.
Even at 133~GeV where the narrow artifact was reported, $\delta f_{\rm sys}\approx 10\%$.\footnote{\url{https://fermi.gsfc.nasa.gov/ssc/data/analysis/LAT_caveats.html}}

To estimate the impact of the unmodeled variation, we add a systematic term to the likelihood function~\citep{Fermi-LAT:2015kyq,DAMPE:2021hsz}
\begin{equation}
    \mathcal{L}(n_{\rm sig}) \to \mathcal{L}(n_{\rm sig}+n_{\rm sys}) \times \frac{1}{\sqrt{2\pi}\sigma_{\rm sys}} \exp\left( -\frac{n^2_{\rm sys}}{2\sigma_{\rm sys}^2}\right),
\end{equation}
where $\mathcal{L}$ is the likelihood function defined in the {\it Methodology} section of the manuscript.
$n_{\rm sig}$ and $n_{\rm sys}$ describe the counts from the signal and the spurious structure, respectively.
The standard deviation of the false signal is $\sigma_{\rm sys}=|\delta f_{\rm sys}|\times b_{\rm eff}$.
The effective background counts are 
\begin{equation}
    b_{\rm eff}=\frac{n_{\rm ph}}{\left(\sum_k \frac{F_{{\rm sig},k}^2}{F_{{\rm bkg},k}}\right)-1},
\end{equation}
where $n_{\rm ph}$ is the total counts in the fit, $F_{{\rm sig},k}$ and $F_{{\rm bkg},k}$ are the binned probability distribution functions for the line and background models at the energy bin $E_{k}$.
To be conservative, the fractional amplitude of the effective area $\delta f_{\rm sys}=10\%$ is chosen.
The same sliding windows analyses as stated previously are performed.

In Fig.~\ref{fig:appx:fsys_ts}, we present the TS values of the line signal considering the variation of the effective area, where the null model is the one with $n_{\rm sig}=0$.
The TS value at 43.2~GeV is 30.5 (21.5) for the first three clusters (the whole sample).
The effective background counts $b_{\rm eff}$ are 26.1 and 54.1 at $\sim 43~\rm GeV$ for the two samples.
Even if the fractional variation is 10\%, the standard deviation of the false signal counts are only 2.6 and 5.4, which is smaller than the statistical fluctuation.
Therefore the significance of the line signal does not change much when considering this type of systematic uncertainty.

\subsection{Verification with Earth limb data}\label{appx:systematics:earthlimb}
\begin{figure}
    \includegraphics[width=0.48\textwidth]{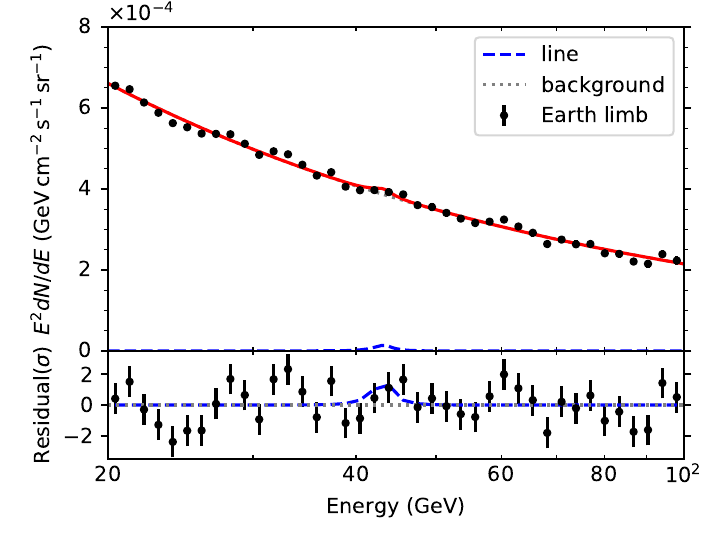}
    \caption{
    The SED of the Earth limb data (upper panel) and the residual in significance (lower panel).
    In the upper panel, the solid line shows the total flux from the best-fit model.
    The blue dashed curve is the signal model, whereas the gray dashed curve is the power-law background.
    The black points in the lower panel give the residual that subtracts the background from the data.
    }
    \label{fig:appx:earthlimb}
\end{figure}

The spectrum of the Earth limb data lacks prominent features, so the data are widely used to test the smoothness of the IRFs~\citep{2013PhRvD..88h2002A,Fermi-LAT:2015kyq,Finkbeiner2013}.
Fig.~\ref{fig:appx:earthlimb} shows the SED of the Earth limb data used in the main text.
The residual that subtracts the power-law background from the data is given in the lower panel.
No line-like excess at 43~GeV is visible.
From the sliding windows analyses, the TS value of the 43~GeV line is only $\sim 0.6$.
The fractional signal at 43~GeV is found to be $f\equiv n_{\rm sig}/b_{\rm eff}\approx 0.02$~\citep{2013PhRvD..88h2002A}.
We also split the Earth limb data into two sets according to the sign of the rocking angle~\citep{Fermi-LAT:2015kyq}.
Neither of them shows any significant line-like feature at around 43~GeV.
The one with the negative rocking angle has a TS value of 1.5 and the fractional signal is $f\approx 0.04$, while the one with the positive rocking angle has a TS value of 0.0.
Considering the low TS value and the fractional signal smaller than $\delta f_{\rm sys}$, no significant structure in the effective area is detected at 43 GeV using the Earth limb data.

\subsection{A summary of the systematic uncertainties}
After a thorough evaluation of the systematic uncertainties, we find:
\begin{itemize}
    \item 
    The 43 GeV line signal exists in different event classes and event types.
    For the event class and event type with the best sensitivity, the TS value is also the largest (Sec.~\ref{appx:systematics:evclass_type}).
    \item 
    The TS value of the line does not drastically change to the size of the energy window.
    It is larger than 20 when we use various window sizes (Sec.~\ref{appx:systematics:winsize}) and background spectral models (Sec.~\ref{appx:systematics:bkgtype}).
    \item 
    No significant line-like artifact around 43~GeV is detected in different incident angles of the null data set (Sec.~\ref{appx:systematics:thetaphi}).
    \item
    No sharp structure in the effective area is found in the Earth limb data (Sec.~\ref{appx:systematics:earthlimb}).
    Even if there is a $\sim 10\%$ peak above the tabulated effective area, it is still hard to explain the 43~GeV line (Sec.~\ref{appx:systematics:fsys}).
\end{itemize}

To summarize, the line is robust against the data selection and background model,
and no significant instrumental systematics are found to explain the line signal plausibly.

\section{Characteristics of the line-like signal}\label{appx:linefeatures}
\begin{figure*}
    \includegraphics[width=0.48\textwidth]{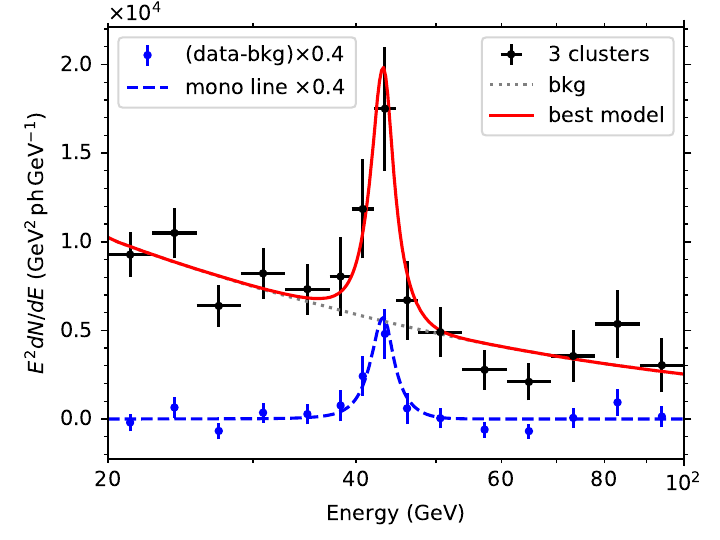}
    \includegraphics[width=0.48\textwidth]{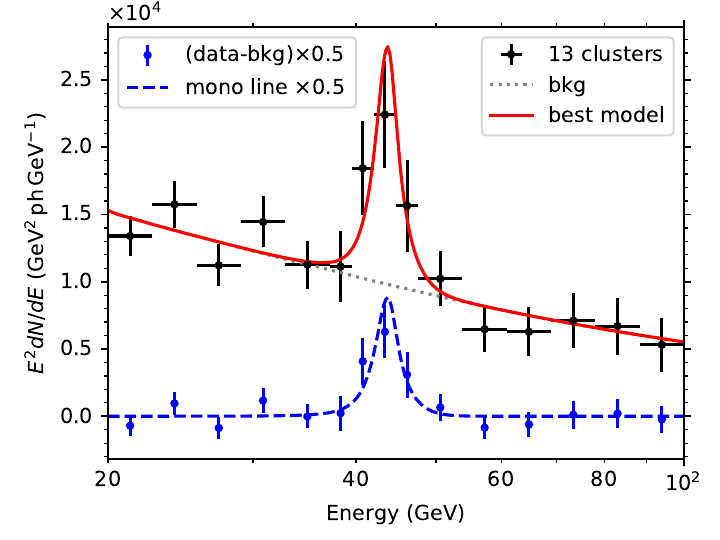}
    \caption{
     The counts spectra of the first three clusters (left panel) and the 13 clusters (right panel).
     The black points are the spectra from the data and the red solid line is the best-fit model using the data from 2~GeV to 100~GeV.
     The blue points are the residual that subtracts the optimal power-law background from the observed counts (gray dotted line).
     The blue dashed curve corresponds to the best monochromatic line model centering at 43~GeV.
    }
    \label{fig:appx:residual}
\end{figure*}

\begin{figure}
    \centering
    \includegraphics[width=0.48\textwidth]{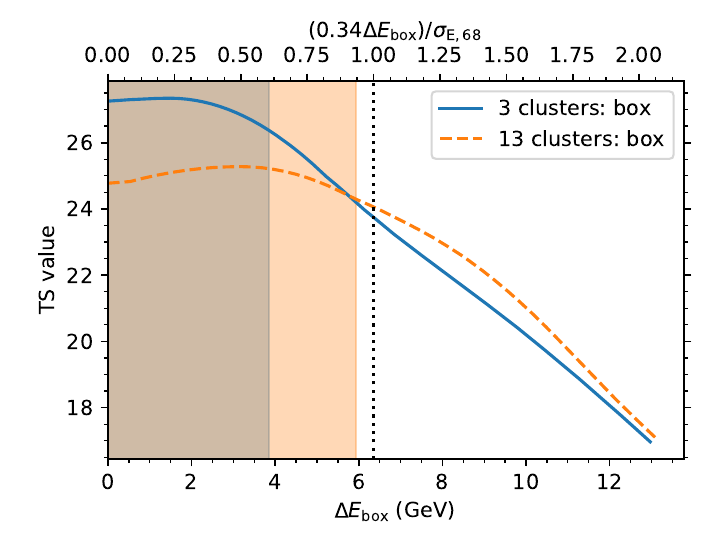}
    \caption{
    The TS value to the energy width of the excess assuming a box-shaped spectrum centering at $\sim 43~\rm GeV$.
    The blue solid line and orange dashed line are the results for the three clusters and 13 clusters respectively.
    The color bands show the $1\sigma$ statistical uncertainties.
    }
    \label{fig:appx:boxwidth}
\end{figure}

\begin{figure}
    \centering
    \includegraphics[width=0.48\textwidth]{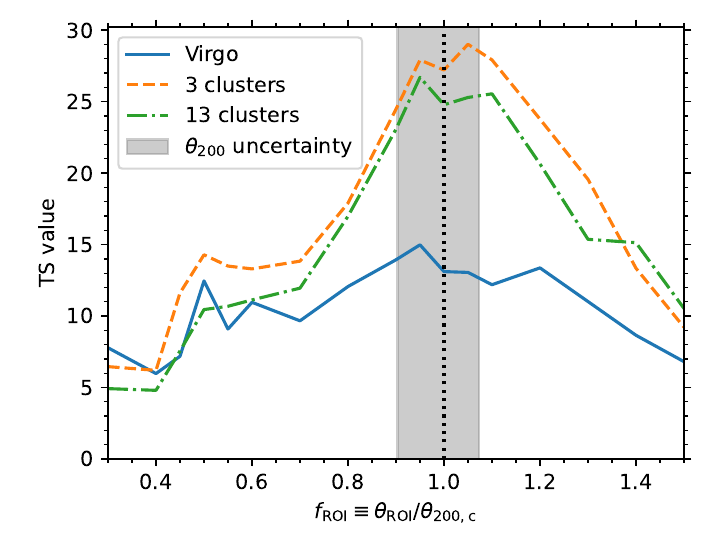}
    \caption{
    The TS value versus the ROI size of the galaxy clusters.
    The ROI radii of all the clusters are defined by $\theta_{\rm ROI}\equiv f_{\rm ROI} \times \theta_{\rm 200,c}$.
    The lines show the results for three samples of galaxy clusters.
    The gray band illustrates the uncertainty of the $\theta_{\rm 200}$ for the first three clusters.
    }
    \label{fig:appx:roisize}
\end{figure}

To further examine the reality of the line signal, we study its spectral and spatial properties in this section.
We adopt the ULTRACLEAN data set with EDISP(1+2+3) event type and restrict the energy from 20 to 100 GeV.\footnote{It is approximately as wide as $\varepsilon=5.0$ in Fig.~\ref{fig:appx:winsize}.}
In Fig.~\ref{fig:appx:residual}, the counts spectra of the two galaxy cluster samples are displayed.
The data are fitted with a monochromatic line and a power-law background, and the best-fit model is given by the red solid line.
The residual counts are calculated by subtracting the background from the observed counts as shown with the blue points.
By visually comparing the residual with the best monochromatic line model (blue dashed line), the excess is well compatible with the line model.

The width of the line is evaluated.
We replace the line model with the following box spectrum
\begin{equation}
    S(E) = N_{\rm s} H(E-E_{-}) H(E_{+}-E)/\Delta E_{\rm box},
\end{equation}
where $H(x)$ is the Heaviside step function.
$E_\pm \equiv E_{\rm c}\pm \Delta E_{\rm box}/2$ are the two energy bounds of the spectrum, where $E_{\rm c}$ is the center energy and $\Delta E_{\rm box}$ is the box width.
The box-shaped spectrum can be produced by the DM annihilation into unstable intermediate particles~\citep{Ibarra2012}.
Besides, the annihilation into the $\gamma Z$ channel, the internal bremsstrahlung~\citep{Bringmann:2012vr}, or even the inverse Compton emission of a cold ultra-relativistic electron cloud in the Klein-Nishina region~\citep{2012A&A...547A.114A} could create a spectral structure slightly broader than the monochromatic line, which can also be approximated using the box spectrum as long as the width not significantly larger than the energy resolution.
The box spectrum is convolved with the energy dispersion function $\bar{D}(E;E'_{\rm line})$ to get the spectrum detected by the Fermi-LAT.
We fit the data from 20 to 100~GeV.
The best-fit center energy for the three clusters (13 clusters) is $43.2\pm0.5~\rm GeV$ ($43.7\pm0.6~\rm GeV$).
Fig.~\ref{fig:appx:boxwidth} shows the TS value versus the box widths given the optimal center energy.
The width $\Delta E_{\rm box}$ is $1.5^{+2.3}_{-1.5}~\rm GeV$ and $3.2^{+2.7}_{-3.2}~\rm GeV$ for the three clusters (the blue solid line) and the 13 clusters samples (the orange dashed line), respectively.
Therefore the intrinsic width of the line-like structure is narrower than the 68\% containment of the energy dispersion ($\sigma_{\rm E,68}\approx 2~\rm GeV$, black dotted line) and is consistent with the monochromatic model.

The asymmetry of the excess is then estimated.
We adopt the bifurcated Gaussian distribution as the signal model
\begin{equation}
    S(E) = 
    \begin{cases}
         N \exp \left(-\frac{(E-E_{\rm c})^2}{2\Delta E_{\rm G,L}^2}\right) & \quad E < E_{\rm c},\\
         N \exp \left(-\frac{(E-E_{\rm c})^2}{2\Delta E_{\rm G,R}^2}\right) & \quad E\geq E_{\rm c},
    \end{cases}
\end{equation}
where $\Delta E_{\rm G,L}$ and $\Delta E_{\rm G,R}$ are the standard deviations on the left and right sides, and $N$ is the prefactor.
This spectral shape can be used to model the $\gamma$-ray signal from the forbidden annihilation or resonance annihilation of the DM particles around black holes~\citep{Cheng:2022esn,Yang:2024jtp}.
To quantify the asymmetry of the signal, we define the ratio $R_{\rm LR}\equiv \Delta E_{\rm G,L}/\Delta E_{\rm G,R}$.
The spectrum is also convolved with energy dispersion functions.
We fix the center energy to 43.2~GeV (3 clusters) and 43.7~GeV (13 clusters), and fit the ratio $R_{\rm LR}$ with the unbinned likelihood analyses.
For three clusters, $R_{\rm LR}=1.1\pm1.6$ and $\Delta E_{\rm G}\equiv \sqrt{\Delta E_{\rm G,L}\Delta E_{\rm G,R}}=0.4\pm7.8~\rm GeV$.
For 13 clusters, $R_{\rm LR}=1.1\pm0.9$ and $\Delta E_{\rm G}=0.91\pm0.87~\rm GeV$.
For both cases, $R_{\rm LR}$ is around 1.0, suggesting no asymmetry in the line-like excess.

The radial size of the signal is also analyzed.
The radii of the ROIs are defined by multiplying the angular size of the virial radii by a common factor $f_{\rm ROI}$: $\theta_{\rm ROI}=f_{\rm ROI}\times \theta_{\rm 200,c}$, where $\theta_{\rm 200,c}$ is the central value of the virial radius.
The overlapped region of Virgo and M49 is only taken into account once.
We perform the unbinned analysis using the data within the ROI radii.
Fig.~\ref{fig:appx:roisize} shows the TS value to the factor $f_{\rm ROI}$ for the Virgo cluster (blue solid line), first three clusters (orange dashed line), and the 13 clusters (green dot-dashed line).
The gray band shows the uncertainty of $\theta_{200}$ for the three clusters, where the boundary is computed with $f_{{\rm ROI},k}=\sqrt{\sum_{i=1}^3{\Omega_{i,k}}/\sum_{i=1}^3{\Omega_{i}}}$ and $k$ labels the lower and upper bounds of $\theta_{200}$. 
The TS values are peaked exactly at $f_{\rm ROI}\sim 1.0$, i.e. the line-like excess mostly exists within the virial radii of the clusters, indicating that the signal may be associated with some physical process in the clusters.

\section{Parameter consistency between the clusters}\label{appx:consistency}
\begin{figure}
    \includegraphics[width=0.48\textwidth]{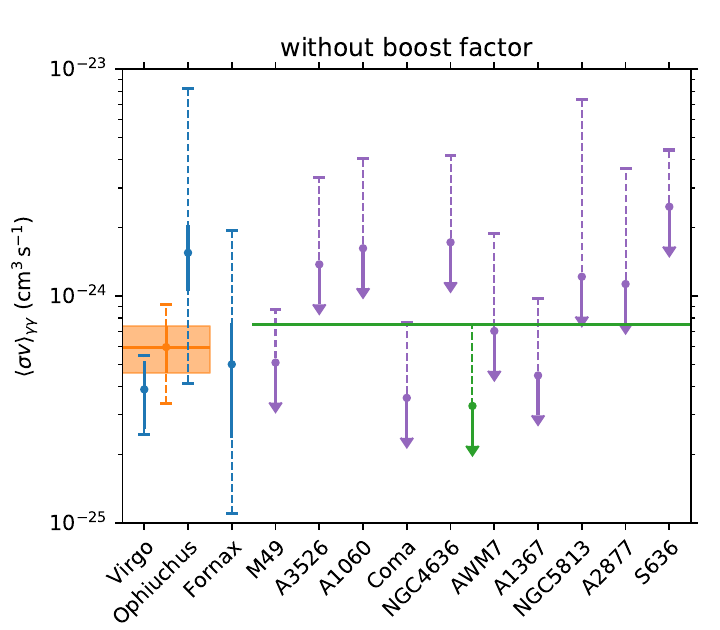}
    \caption{
        The annihilation cross section $\left< \sigma v \right>_{\gamma\gamma}$ for each cluster from the sliding window analysis.
        For the first three clusters, we present the best-fit values (blue points) with the $1\sigma$ statistical uncertainties (blue solid error bars).
        For the others, the 95\% upper limits are presented with the purple solid arrows.
        The thin dashed lines illustrate the systematic uncertainties from the J-factors.
        The results for Virgo and Ophiuchus are presented with the orange rectangle, while the constraints from the last 10 clusters are given with green upper limits.
    }
    \label{fig:appx:sv_srcs}
\end{figure}

If the line signal is indeed from the DM annihilation, the annihilation cross sections $\left< \sigma v \right>_{\gamma\gamma}$ from each source should be consistent.
Fig.~\ref{fig:appx:sv_srcs} presents the constraints/best-fit results of the annihilation cross section.
We can see that when considering the statistical uncertainties from the $\gamma$-ray data analysis (solid error bars), the results between different clusters are basically consistent, with only the Ophiuchus and Coma clusters deviating from the overall fit (note also that, according to the statistical law, on average, $\sim4$ out of every 13 measurements are expected to deviate by $>1 \sigma$).
Moreover, these results are based on fixed J-factor values.
In fact, the measurements of the J-factors are far from perfect.
There are large uncertainties in the measurements of galaxy cluster radii and masses, meaning the J-factor values themselves carry uncertainties.
When considering these J-factor uncertainties, the dark matter parameters derived from these galaxy clusters become even more consistent between sources, as indicated by the dashed error bars in the figure.

\section{Upper limits of annihilation cross section}\label{appx:containment}

\begin{figure*}
    \includegraphics[width=0.48\textwidth]{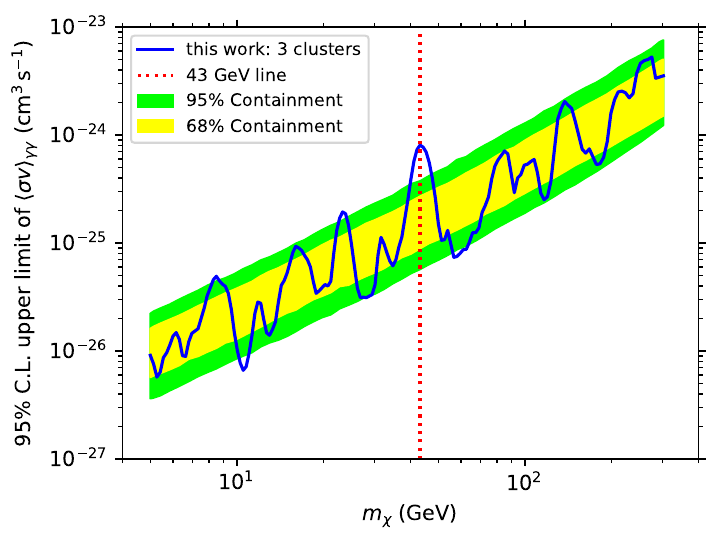}
    \includegraphics[width=0.48\textwidth]{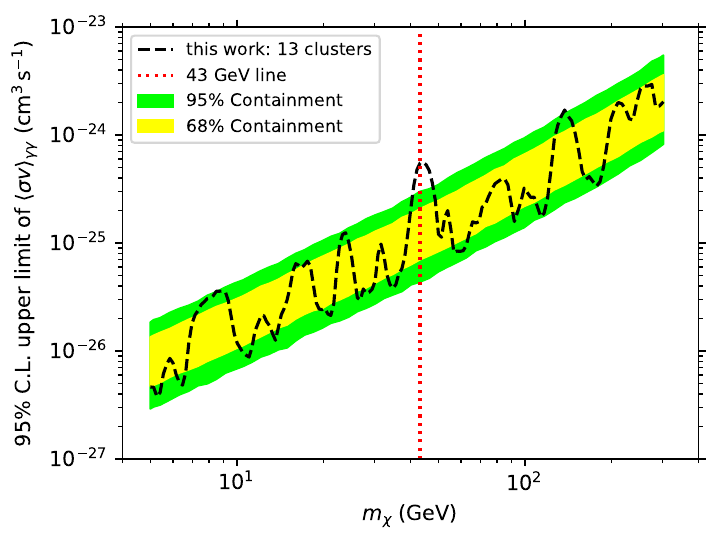}
    \caption{
    The 95\% confidence level $\left< \sigma v \right>_{\gamma\gamma}$ upper limits for the top three clusters (left) and the 13 clusters (right).
    The yellow (green) bands show the 68\% (95\%) expected containment bands derived from the 3000 simulations of the best-fit no-DM models parameterized with the LogParabola spectrum.
    The red dotted line shows the energy of $\sim 43~\rm GeV$.
    Please note that the boost factors are not accounted for in the upper limits.
    }
    \label{fig:appx:containment}
\end{figure*}

Fig.~\ref{fig:appx:containment} gives the 95\% C.L. upper limits of $\left< \sigma v \right>_{\gamma\gamma}$ for the top three clusters (left panel) and the total 13 clusters (right panel).
The center values of the J-factors in Tab.~\ref{tab:appx:clusters} are adopted to convert the upper limits of the line normalization to those of the cross sections.
Since the boost factors are not considered in the calculation, the constraints are conservative.

To illustrate the statistical uncertainties of the constraints, we make the containment bands obtained by 3000 MC simulations.
The LogParabola models fitted from the stacked data between 2.5 and 450~GeV (see Sec.~\ref{appx:systematics:bkgtype}) are chosen as the input of the simulations.
In each simulation, we simulate the photon events based on the spectral models and conduct the same analysis procedure as stated above.
The yellow and green bands in Fig.~\ref{fig:appx:containment} correspond to the 68\% and 95\% expected containment regions of the DM constraints, respectively.
The fluctuations of the constraints are mostly located within the 95\% regions except that at $\sim 43~\rm GeV$ (the red dotted line).

In the baseline results, we do not account for the uncertainties of the J-factors.
To check their impact, we rewrite the signal model $F_{\rm s}(E)\bar{\epsilon}(E'_{\rm line})$ in Eq.(\ref{eq:lnLsig}) using ${\tilde N}_{\rm s} \sum_{i=1}^{N_{\rm gcl}} D_i(E;E'_{\rm line}) \epsilon_i(E'_{\rm line}) J_i$, where $\epsilon_i$ is the exposure, $D_i$ is the energy dispersion function, and $J_i$ is the J-factor for the $i$-th cluster.
${\tilde N}_{\rm s}$ is the normalization.
The J-factor is assumed to follow the log-normal distribution~\citep{Fermi-LAT:2015att,CTA2024_line}
\begin{equation}
    {\rm Log}\mathcal{N}(\lg J) \propto \frac{1}{J_{\rm c} \sigma_{\lg J}}\exp\left[ -\left(\frac{\lg J - \lg J_{\rm c}}{\sqrt{2}\sigma_{\lg J}}\right)^2 \right],
\end{equation}
where $\lg J_{\rm c}$ and $\sigma_{\lg J}$ are the central value and uncertainty of the J-factor, respectively.
Then the total likelihood with J-factor uncertainties can be written as
\begin{equation}\label{eq::llk_stack_prior}
\begin{aligned}
    \tilde{\mathcal{L}}({\tilde N}_{\rm s},E_{\rm line},\Theta_{\rm b},\lg {\bm J}) &= \left[\prod_{i=1}^{N_{\rm gcl}} {\rm Log}\mathcal{N}(\lg J_i)\right] \\
     &\times  \mathcal{L}({\tilde N}_{\rm s},E_{\rm line},\Theta_{\rm b},\lg {\bm J}),
\end{aligned}
\end{equation}
where $\mathcal{L}$ is the baseline likelihood for null and signal models.
We perform the same sliding windows analysis and find that the TS values are the same as those in Fig.~\ref{fig:SED}.
It is reasonable since the J-factors are highly correlated with each other in the likelihood and can be absorbed by the normalization ${\tilde N}_{\rm s}$ no matter what the values are.
The DM constraints are slightly weakened, because the J-factors converge to the optimal values of the log-normal distribution $\lg \hat{J} = \lg J_{{\rm c}}-\ln(10)\sigma^2_{\lg J}$, smaller than the central values $\lg J_{{\rm c}}$.
According to the fittings, we find that the upper limits for the 13 (3) clusters are a factor of $\approx \sum_{i=1}^{N_{\rm gcl}} J_{{\rm c},i}/\sum_{i=1}^{N_{\rm gcl}} \hat{J}_i = 1.56$ ($1.36$) higher than the baseline results (Fig.~\ref{fig:appx:containment} and Fig.~\ref{fig:Constraint}).

\begin{figure*}
    \includegraphics[width=0.48\textwidth]{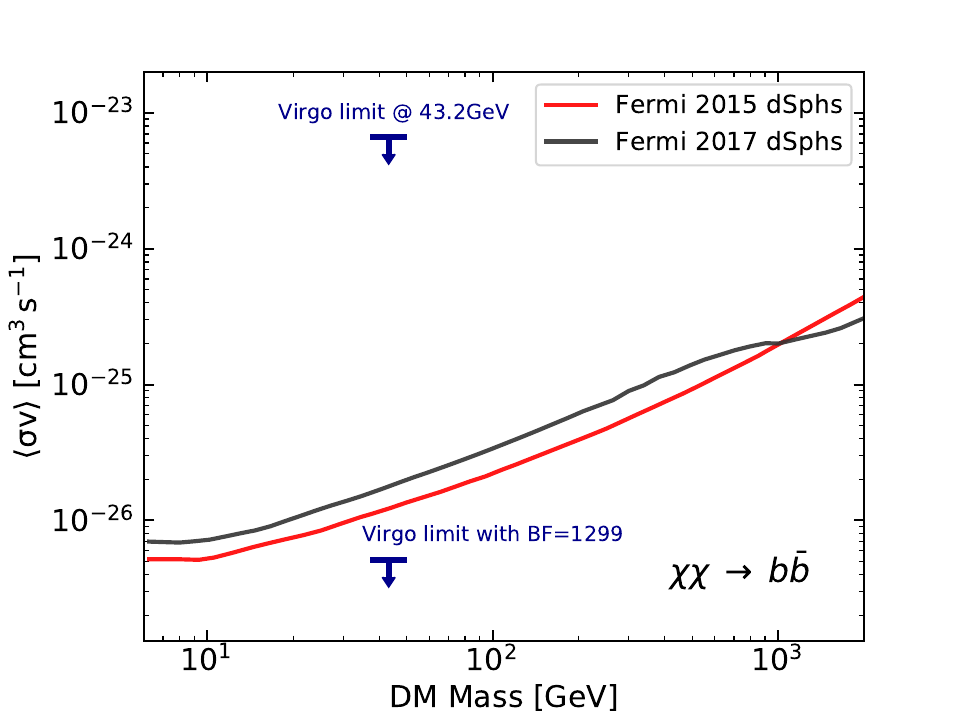}
    \includegraphics[width=0.48\textwidth]{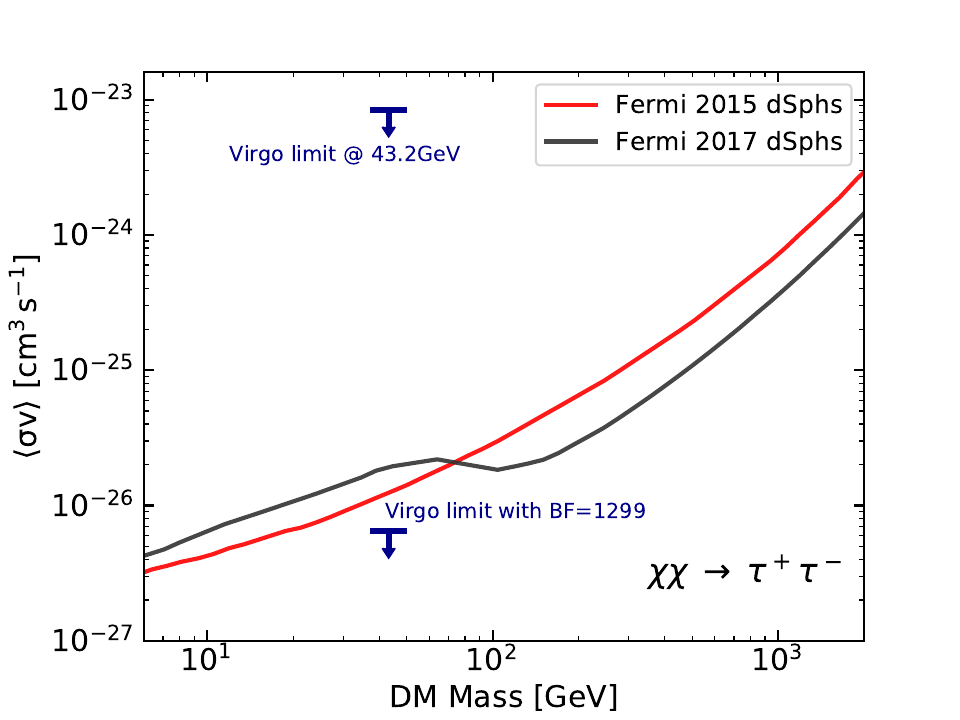}
    \caption{
    The constraints from the continuum emission from the DM annihilation to quarks or charged leptons.
    We search for such emission toward the Virgo cluster and find no signal. Upper limits on $\left<\sigma v\right>$ at a DM mass of 43.2 GeV are therefore placed. We also plot the upper limits assuming an optimistic boost factor of 1299 for Virgo.}
    \label{fig:appx:continuum}
\end{figure*}

\section{Search for the associated continuum emission}\label{appx:continuum}
Theoretically, dark matter annihilation to photons is loop-suppressed, and it is expected to predominantly annihilate through quark and charged-lepton channels. Therefore, if the $\sim43$ GeV $\gamma-$ray line signal analyzed in this paper is indeed produced by dark matter, it would be accompanied by a continuum $\gamma$-ray component generated from dark-matter annihilation to quarks or charged leptons.
Here we attempt to search for such a continuous spectrum signal. Our analysis primarily targets at the Virgo cluster, which has the highest expected J-factor. We performed a standard Fermi-LAT binned likelihood analysis using a $20^\circ\times20^\circ$ ROI with a bin size of 0.1$^\circ$. 
The background model includes all 4FGL-DR4 sources \cite{Fermi-LAT:2022byn} and two large-scale diffuse components (the Galactic interstellar emission and isotropic diffuse component). We add a dark matter component into the background model to examine the existence of the continuum emission, with its spatial template derived from considering an NFW density distribution. The related parameters $r_s$ and $\rho_0$ are given by $M_{200}$ and $R_{200}$ listed in Table~\ref{tab:appx:clusters}. We consider two representative annihilation channels, $b\bar{b}$ and $\tau^+\tau^-$, with the corresponding annihilation spectra generated with PPPC4DMID \cite{Cirelli:2010xx}.

For our purpose, we fix the DM mass at {43.2 GeV}, the best-fit mass of the top three clusters in Fig.~\ref{fig:Constraint} of the main text. We do not detect any significant continuum $\gamma$-ray emission beyond the background of dark-matter annihilation for both the $b\bar{b}$ and $\tau^+\tau^-$ channels, with the TS values approximately 0. Therefore, we place the upper limits on the annihilation cross section $\left<\sigma v\right>$ of dark matter to these two final states. 
The results are shown in Fig.~\ref{fig:appx:continuum}. 
For comparison, also plotted in the figure are the constraints given by Fermi-LAT observations of dwarf spheroidal galaxies \cite{Fermi-LAT:2015att,Fermi-LAT:2016uux}, which generally represent the strongest limits on the dark matter annihilation cross-section in this mass range.
We have also plotted the upper limits that assumed an optimistic boost factor of 1299 (such a large boost factor is probably necessary to ensure that the $\sim43$~GeV line signal does not conflict with the line search results from the observations of the Galactic center region \cite{DAMPE:2021hsz,Cheng:2023chi}). It can be seen that under this choice of boost factor, the dark matter cross-section is constrained to a very low value, which may impose challenges for model building if the line signal is true.

\section{The DAMPE data analysis}\label{appx:dampe}
\begin{figure}
    \includegraphics[width=0.48\textwidth]{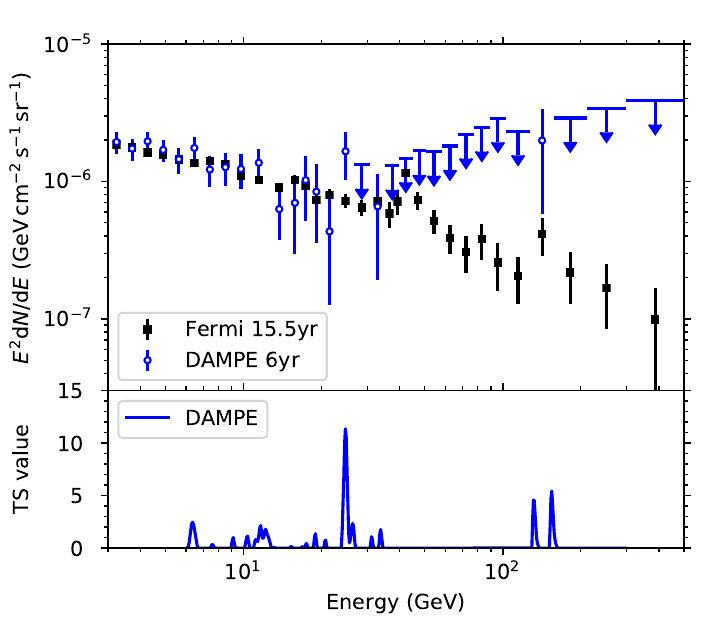}
    \caption{
        The upper panel demonstrates the stacked SED of the 13 galaxy clusters observed by DAMPE (the blue points) and Fermi-LAT (the black points).
        Upper limits are presented if the photon count within the energy bin is less than two.
        The lower panel shows the TS values of $\gamma$-ray lines from the sliding window analyses of the DAMPE data.
    }
    \label{fig:appx:dampe_flux}
\end{figure}

DAMPE is a space-borne pair-conversion $\gamma$-ray telescope covering the energy range from $\sim 2~\rm GeV$ to $\sim 10~\rm TeV$~\citep{DAMPE:2017cev,DAMPE:2021hsz,DAMPE2025a}.
It has an excellent energy resolution: at $43~\rm GeV$, the energy resolution is $\Delta E/E \approx 1\%$ with the total acceptance of $1800~\rm cm^2\,sr$~\citep{DAMPE:2021hsz}.
Utilizing the publicly available six years (\url{https://dampe.nssdc.ac.cn/dampe/mission.php}) of High-Energy-Trigger DAMPE photon data~\citep{Xu2018,DAMPEdata2023}, we searched for the line signal in the sample of 13 galaxy clusters.
The latest {\tt DmpST} package~\citep{Duan2019}, which incorporates the latest calibrations on the instrumental response functions~\citep{Ambrosi2019,Shen2024,Duan2025}, is employed in the analysis.

We selected photons with energy $\geq 3~\rm GeV$ from the 13 clusters within the radii of $\theta_{\rm ROI,0}$ and calculated the stacked SED.
In the calculation, we considered the $\phi$-dependence of the effective area due to the nonuniform exposure.
As depicted in the upper panel of Figure~\ref{fig:appx:dampe_flux}, the flux from the DAMPE (blue points) is well consistent with that of the Fermi-LAT (black points).
Owing to the small acceptance of DAMPE, no photon is detected between 34~GeV and $130~\rm GeV$ and the flux upper limits are drawn.

The unbinned sliding windows analysis was adopted to search for the line signal above 6~GeV.
No significant $\gamma$-ray lines are found by DAMPE as shown in the lower panel of Figure~\ref{fig:appx:dampe_flux}, especially the TS value at $\sim 43~\rm GeV$ is zero.
Based on the best-fit parameters in the Fermi-LAT sliding windows analysis, the expected photon number of the line is found to be 1.53.
It is 21.6\% in probability for zero detected photon within the energy bin of $42-44~\rm GeV$.
Therefore, the null result based on the DAMPE data is still consistent with the result of Fermi-LAT.

\section{Prospect of the detection with Cherenkov Telescope Array}\label{appx:CTA}
\begin{figure*}[!t]
    \includegraphics[width=0.4\textwidth]{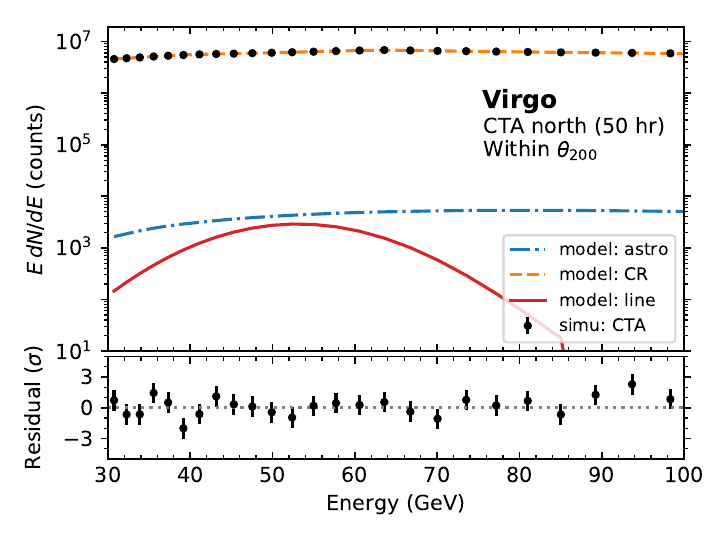}
    \includegraphics[width=0.4\textwidth]{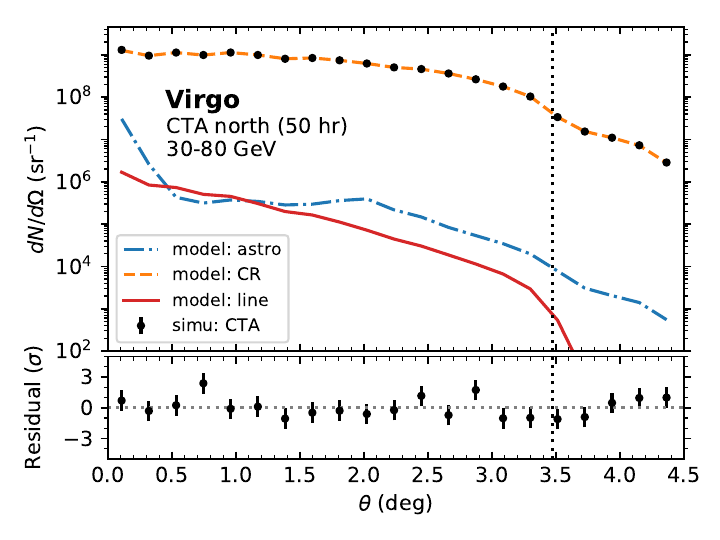}
    \includegraphics[width=0.4\textwidth]{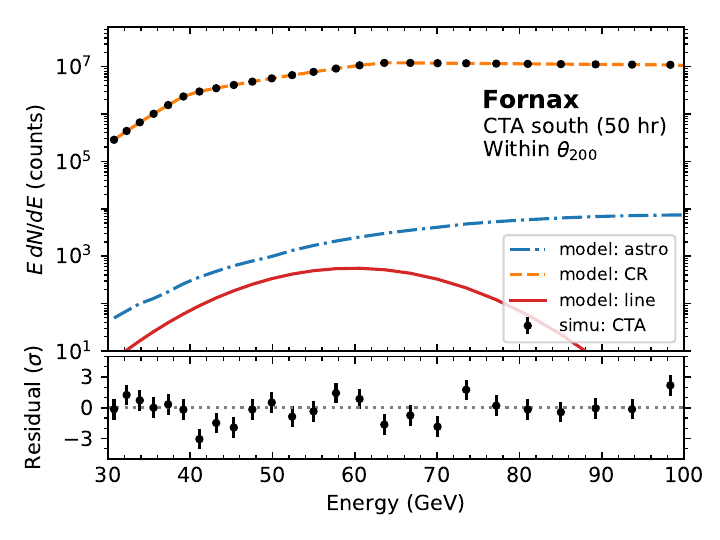}
    \includegraphics[width=0.4\textwidth]{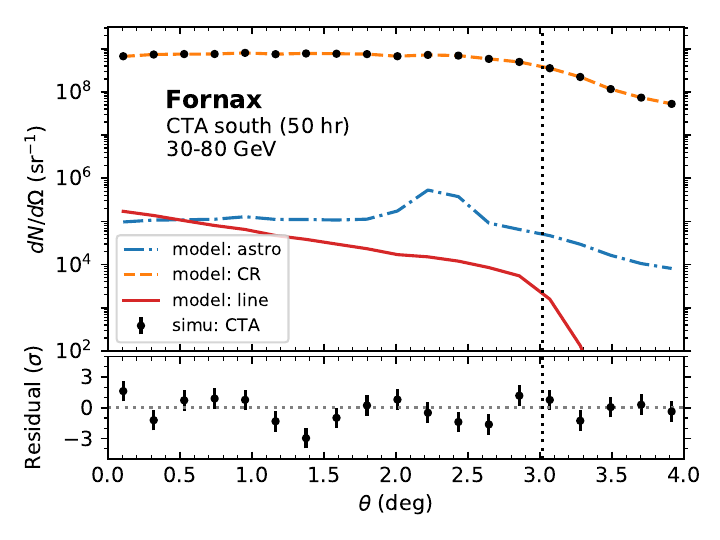}
    \includegraphics[width=0.4\textwidth]{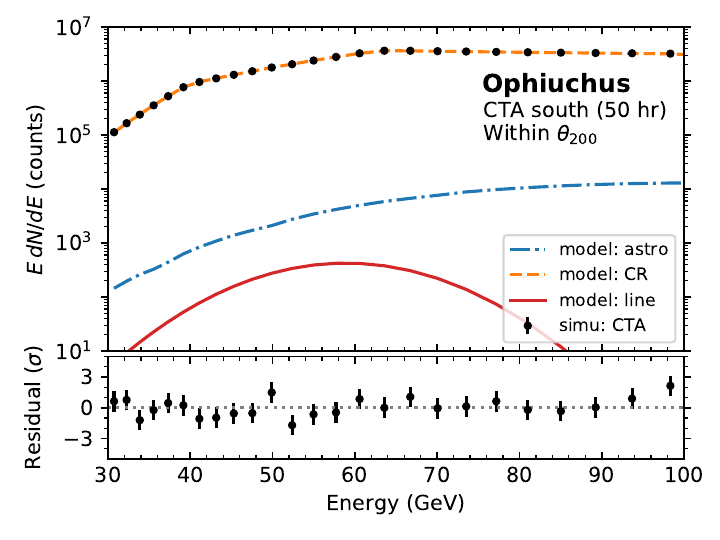}
    \includegraphics[width=0.4\textwidth]{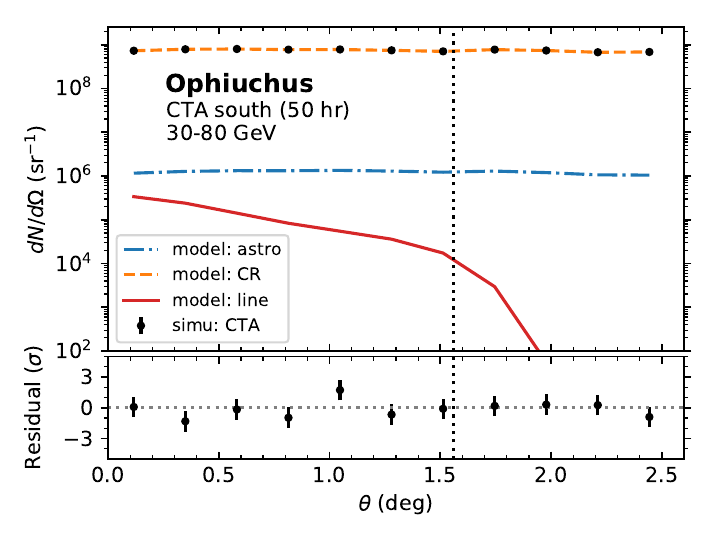}
    \caption{
        The counts spectra (left panels) and angular distribution (right panels) for the 50-hr mock data of the Virgo (top), Fornax (middle), and Ophiuchus (bottom) clusters.
        In the upper part of each figure, the red solid line, blue dot-dashed line, and orange dashed line are the predicted spectra for the 43~GeV line, the astrophysical components, and the instrumental CR background.
        The instrumental responses are considered for all the components.
        The black points in the lower part of each figure correspond to the residual for the model without the line.
        The black dotted line in the right panels shows the angular size of the clusters.
    }\label{fig:appx:cta_counts}
\end{figure*}

Cherenkov Telescope Array Observatory (CTAO) is the next-generation ground-based $\gamma$-ray observatory~\citep{CTAConsortium:2013ofs}.
It is made of three types of telescopes: the Large-Sized Telescope (LST) for $20-150~\rm GeV$ range, the Medium-Sized Telescope (MST) for $0.15-5~\rm TeV$ range, and the Small-Sized Telescope (SST) for $5-300~\rm TeV$ range.
CTAO has two observation sites each focusing on one hemisphere: the southern one in Paranal, Chile, and the northern one in La Palma, Spain.
In the ``Alpha'' configuration, the northern array has 4 LSTs and 9 MSTs, while the southern array has 14 MSTs and 37 SSTs.
The effective areas for northern and southern arrays at $43~\rm GeV$ are approximately $2\times10^4~\rm m^2$ and $3\times10^3~\rm m^2$, respectively.\footnote{\url{https://www.ctao.org/for-scientists/performance/}}
The energy resolution is $\sim 15\%$ below 100~GeV.
Thanks to the large effective area, CTA will be a powerful instrument for dark matter indirect detection~\citep{CTA2019_science,CTA2021_GC,CTA2024_line,Rodd2024_TeVline}.
In this section, we focus on the prospect of the line detection in the Virgo, Fornax, and Ophiuchus clusters.

We simulate the events around the clusters using the software {\tt ctools v2.1.0}~\citep{Knodlseder2016} and the responses {\tt prod5 v0.1}.\footnote{\url{https://doi.org/10.5281/zenodo.5499840}}
The Virgo cluster is targeted by the northern array, while the Fornax and Ophiuchus clusters are targeted by the southern one.
We consider the following astrophysical components: the Galactic diffuse emission (GDE) model {\tt gll\_iem\_v07}~\citep{Fermi2016_GDE}, the isotropic diffuse $\gamma$-ray background (model A in~\citep{Fermi2015_IGRB}), and the bright point sources (PS) in Fermi-LAT 4FGL-DR4 catalog~\citep{Fermi2023_4FGLDR4}.
The cosmic ray (CR) induced instrumental background {\tt CTAIrfBackground} is also added to the model.
For the line component, we adopt a narrow Gaussian as the spectral model and the normalized J-factor map with substructures as the spatial model.
The best-fit DM parameters ($m_\chi=43.7~\rm GeV$, $\left<\sigma v \right>_{\gamma\gamma}=3.8\times10^{-25}~\rm cm^3\,s^{-1}$) are adopted as the input spectral parameters.
The J-factor template of the line consists of the two components:
one is from the main halo and can be calculated with $J_{\rm main}(\theta)=\int_{\rm los} \rho_{\rm NFW}^2 (r(s,\theta)) {\rm d}s$;
the other comes from the DM annihilation in the subhalos, whose profile within the angular size of the cluster $\theta_{200}$ can be modelled with~\cite{Gao:2011rf}
\begin{equation}
    J_{\rm sub}(\theta) = \frac{b_{\rm sh} J_{\rm NFW}}{\pi \ln(17)}\frac{1}{\theta_{200}^2 + 16 \theta^2},
\end{equation}
where $J_{\rm NFW}$ and $b_{\rm sh}$ are the J-factor and boost factor, respectively, which are listed in Tab.~\ref{tab:appx:clusters}.
We make the spatial template for the line ($201\times 201$ pixels with $0.05^\circ$ pixel width) by adding the two components together, i.e. $J_{\rm tot}(\theta)=J_{\rm main}(\theta)+J_{\rm sub}(\theta)$, and then normalizing the solid angle integration of the map to one.
We simulate the events for 50-hr exposure time assuming the zenith angle of $20^\circ$ with {\tt ctobssim}.\footnote{\url{http://cta.irap.omp.eu/ctools/users/tutorials/cta/quickstart/simulating.html}}
Please note that only events more energetic than 30~GeV can be simulated due to the limitation of the IRFs.
The left panels of Fig.~\ref{fig:appx:cta_counts} display the spectra of the mock data within the virial radii of the clusters,\footnote{
Since the southern array is not equipped with LSTs, fewer counts are collected below $\sim 70~\rm GeV$ for the Fornax and Ophiuchus clusters.}
whereas the right panels present the distribution of the mock data with respect to the angular distance to the cluster center.
The 43~GeV line only contributes $\sim 100-1000$ photons compared to the total events counts of $\sim 10^6-10^7$.
Even though the total counts from the line are comparable to the astrophysical emission, they are much fewer than the cosmic-ray background.

We estimate the significance of the line signal using the binned likelihood analysis~\citep{CTA2024_line}.
We define the radius of ROI with $\theta_{\rm ROI,CTA} \equiv \theta_{200}+1^\circ$, and select the events inside the region.
The ROI radius is slightly larger than the cluster size in order to better account for the background emission.
The mock data from 30~GeV to 200~GeV are split into a counts cube with 40 logarithmic energy bins and $100\times 100$ pixels with $0.1^\circ$ width using {\tt ctbin}.
We calculate the expected counts cubes for the astrophysical components ($\mu_{\rm ast} \equiv \mu_{\rm gde}+\mu_{\rm iso}+\mu_{\rm ps}$), $\gamma$-ray line ($\mu_{\rm sig}$) and CR background ($\mu_{\rm cr}$) within the ROI using {\tt ctmodel}.
The total predicted counts in the pixels with the central
energy of $E$ and coordinate of $\bm x$ are
\begin{equation}\label{eq::appx:CTAO_counts}
    \mu_{\rm tot}({\bm x},E) = \mu_{\rm cr} + b_{\rm sig} \mu_{\rm sig} + b_{\rm ast}({E}/{E_0})^{\gamma_{\rm ast}} \mu_{\rm ast},
\end{equation}
where $b_{\rm sig}$ is the normalization factor for the line.
$b_{\rm ast}$ and $\gamma_{\rm ast}$ are the prefactor and index for the scale of the astrophysical emission, respectively.
$E_0 = 100~\rm GeV$.
The binned Poisson likelihood function is adopted
\begin{equation}
    \ln \mathcal{L}_{\rm cta}(b_{\rm sig},b_{\rm ast}, \gamma_{\rm ast})\equiv \sum_{ij} [n_{ij} \ln (\mu_{{\rm tot},ij}) - \mu_{{\rm tot},ij}].
\end{equation}
$n_{ij}$ and $\mu_{{\rm tot},ij}\equiv\mu_{\rm tot}({\bm x}_j, E_i)$ are the mock data counts and the expected counts in the pixel centering at the energy $E_i$ and the coordinate ${\bm x}_j$.
The TS value is defined as ${\rm TS} \equiv -2\ln  [\mathcal{L}_{\rm cta}(0, \hat{b}_{\rm ast,0},\hat{\gamma}_{\rm ast,0})/\mathcal{L}_{\rm cta}(\hat{b}_{\rm sig,1}, \hat{b}_{\rm ast,1},\hat{\gamma}_{\rm ast,1}) ]$,
where the quantity with the hat is the best-fit value.
The subscripts 0 and 1 represent the null and alternative models, respectively.
The TS values of the line for Fornax and Ophiuchus clusters are close to zero (${\rm TS}<0.01$).
For the Virgo cluster, however, a weak excess with ${\rm TS}\approx 1.0$ is expected.
Notably, the systematic uncertainty from the correlated noise is not considered in our analysis~\citep{CTA2024_line,CTA2021_GC}, so the TS value in observations may be even smaller.
To conclude, it is challenging to detect the 43~GeV line with CTAO.

\end{document}